\documentclass[a4paper]{article}

\usepackage[]{authblk}
\usepackage{anysize}
\marginsize{2cm}{2cm}{2cm}{2cm}

\def\degr{\hbox{$^\circ$}}

\usepackage{graphicx}
\usepackage{txfonts}
\usepackage{natbib}
\usepackage{longtable}

\begin{document}

	\title{Design, development and verification of the 30 and 44 GHz front-end modules for the Planck Low Frequency Instrument}

	\author[1]{Davis, R.J.}
	\author[1]{Wilkinson A.}
	\author[1]{Davies, R.D.}
	\author[1]{Winder, W.F.}
	\author[1]{Roddis, N.}
	\author[1]{Blackhurst, E.J.}
	\author[1]{Lawson, D.}
	\author[1]{Lowe, S.R.}
	\author[1]{Baines, C.}
	\author[1]{Butlin, M.}
	\author[1]{Galtress, A.}
	\author[1]{Shepherd, D.}
	\author[2]{Aja, B.}
	\author[2]{Artal, E.}
	\author[3]{Bersanelli, M.}
	\author[4]{Butler, R.C.}
	\author[5]{Castelli, C.}
	\author[4]{Cuttaia, F.}
	\author[6]{D'Arcangelo, O.}
	\author[7]{Gaier, T.}
	\author[8]{Hoyland, R.}
	\author[9]{Kettle, D.}
	\author[10]{Leonardi, R.}
	\author[4]{Mandolesi, N.}
	\author[3]{Mennella, A.}
	\author[10]{Meinhold, P.}
	\author[11]{Pospieszalski, M.}
	\author[4]{Stringhetti, L.}
	\author[3]{Tomasi, M.}
	\author[4]{Valenziano, L.}
	\author[3]{Zonca, A.}
	\affil[1]{Jodrell Bank Centre for Astrophysics, Alan Turing Building, The University of Manchester, Manchester, M13 9PL, UK}
	\affil[2]{Departamento de Ingenier\'{i}a de Comunicaciones, Universidad de Cantabria, Avenida de los Castros s/n. 39005 Santander, Spain}
	\affil[3]{Physics Department, Universit\`{a} degli Studi di Milano, Via Celoria 16, 20133 Milano, Italy}
	\affil[4]{INAF IASF Bologna, Via Gobetti,101, 40129, Bologna, Italy}
	\affil[5]{Science and Technology Facilities Council, Swindon, Wiltshire, SN2 1SZ, UK}
	\affil[6]{IFP-CNR, Via Cozzi 53, Milano, Italy}
	\affil[7]{Jet Propulsion Laboratory, Pasadena, California, USA}
	\affil[8]{Instituto de Astrof\'{i}sica de Canarias, C/ V\'{i}a L\'{a}ctea, s/n E38205 - La Laguna (Tenerife), Spain}
	\affil[9]{School of Electrical and Electronic Engineering, The University of Manchester, Manchester, M60 1QD, UK}
	\affil[10]{Department of Physics, University of California, Santa Barbara, CA 93106-9530, USA}
	\affil[11]{National Radio Astronomy Observatory, 520 Edgemont Rd, Charlottesville, VA 22903-2475 USA}

	\date{published in JINST, 28 Dec, 2009}

	\maketitle

	\begin{abstract}
		We give a description of the design, construction and testing of the 30 and 44 GHz Front End Modules (FEMs) for the Low Frequency Instrument  (LFI) of the Planck mission to be launched in 2009.  The scientific requirements of the mission determine the performance parameters to be met by the FEMs, including their linear polarization characteristics.

		The FEM design is that of a differential pseudo-correlation radiometer in which the signal from the sky is compared with a 4-K blackbody load.  The Low Noise Amplifier (LNA) at the heart of the FEM is based on indium phosphide High Electron Mobility Transistors (HEMTs).  The radiometer incorporates a novel phase-switch design which gives excellent amplitude and phase match across the band.

		The noise temperature requirements are met within the measurement errors at the two frequencies.  For the most sensitive LNAs, the noise temperature at the band centre is 3 and 5 times the quantum limit at 30 and 44 GHz respectively. For some of the FEMs, the noise temperature is still falling as the ambient temperature is reduced to 20 K.  Stability tests of the FEMs, including a measurement of the 1/f knee frequency, also meet mission requirements.

		The 30 and 44 GHz FEMs have met or bettered the mission requirements in all critical aspects.  The most sensitive LNAs have reached new limits of noise temperature for HEMTs at their band centres.  The FEMs have well-defined linear polarization characteristcs.
	\end{abstract}

	\section{Introduction}

		Observations of the Cosmic Microwave Background (CMB) provide unique information about the early history of the Universe some 380,000 years after the Big Bang.  The CMB power spectrum as a function of the angular frequency, \textit{l}, gives critical data for cosmology.  Up to the present, the angular frequency  coverage is obtained from a combination of ground and balloon (high-\textit{l}) observations of  restricted areas and space (low-\textit{l}) all-sky observations. Ground-based instruments sampling high-\textit{l} values include CBI \citep[][]{2004ApJ...609..498R}, the VSA \citep[][]{2004MNRAS.353..732D}, DASI \citep[][]{2002Natur.420..772K}, QUaD \citep[][]{2004SPIE.5498..396C} and ACBAR \citep[][]{2004ApJ...600...32K}; balloon instruments include BOOMERANG \citep[][]{2006ApJ...647..823J} and Archeops \citep[][]{2007A&A...467.1313M}. The space-based instruments giving all-sky coverage are COBE \citep[][]{1990ApJ...360..685S} and WMAP \citep[][]{2003ApJS..145..413J,2007ApJS..170..263J,2007ApJS..170..288H}. The Planck satellite will give full sky coverage over a wide \textit{l} range in a single instrument with data collected at all frequencies simultaneously.

		From the earliest days of CMB research it was recognized that Galactic foregrounds were a significant contaminant of the CMB.  At the lower frequencies, synchrotron \citep[e.g.][]{1982A&AS...47....1H} and free-free emissions \citep[][]{2003MNRAS.341..369D} were well-established foregrounds but their accuracy is insufficient for precise measurement of the CMB and as indicated below, Planck will be used to improve this.  A spinning dust (``anomalous emission'') component \citep[][]{1998ApJ...508..157D,1999ApJ...527L...9D} was later identified; it dominated the foregrounds in the frequency range 15 to 40 GHz \citep[][]{2003MNRAS.345..897B,2006MNRAS.370.1125D}. At frequencies above $\sim$100~GHz, the dominant foreground is the thermal (vibrational) emission from interstellar dust which has a range of temperatures and particle sizes \citep[][]{1990A&A...237..215D}. HFI will enable an accurate measurement of this foreground.

		To derive the best estimate of the CMB sky distribution, it is necessary to have a wide frequency coverage in the Planck armoury in order to identify and remove the Galactic foregrounds.  The Planck satellite covers the frequency range 30 to 857~GHz with two technologies.  The Low Frequency Instrument (LFI) consists of low noise amplifiers at 30, 44 and 70~GHz.  The High Frequency Instrument (HFI) uses bolometers at 100, 143, 217, 353, 545 and 857~GHz. The frequency bands which are least contaminated by foregrounds, and hence best for observing the CMB,  are those in the octave either side of 100~GHz.  The 70~GHz channel is the channel where foregrounds are minimum both in intensity and linear polarization.

		This wide frequency coverage in a deep all-sky survey will provide crucial information on all the Galactic components. The 30 and 44~GHz channels are particularly important in determining the properties of the synchrotron, free-free and spinning dust components when combined with new ground-based surveys at lower frequencies.  

		All the receivers/detector bands on Planck between 30 and $353$~GHz will be sensitive to linear polarization.  Special attention has been paid to the purity of the polarization detection system since the E-mode polarization of the CMB is weak ($\sim$~2~$\mu$K) and any B-mode polarization will be weaker by factors of 10 or more. The linear polarization of the synchrotron emission is relatively strong, so accurate measurements over a wide frequency range will contribute substantially to an understanding of the structure of the Galactic magnetic field.  Careful measurements of the polarization in all the frequency channels are necessary to detect and quantify the polarization of the free-free and dust -- both spinning and thermal. 

		In this paper we describe the design, development and testing of the cryogenic section (the front end modules, FEMs) of the LFI radiometers at 30 and 44~GHz. The structure of the paper is as follows. Section \ref{specification} describes the specification and rationale of the radiometer chain at 30 and 44~GHz, along with a description of the input structure (the horns) and the back end modules to which the FEMs are attached.  Section \ref{design} describes the design of the various components of the FEMs. Section \ref{assembly} gives the details of the FEM assembly and qualification. The radiometric performance of the FEMs is given in Section \ref{performanceresults}.  An overall summary of the expected performance of the 30 and 44~GHz radiometers in orbit is provided in Section \ref{conclusions}.

\section{Specification of the Radiometer chain at 30 and 44~GHz}
\label{specification}

	The principal observational objective of Planck is to produce maps of the whole sky in the 9 frequency channels listed in Table 1. The telescope is an off-axis aplanatic design with a diameter of 1.5 m. The spacecraft will spin at $\sim$~1 rpm around an axis offset by $\sim$~85\degr from the telescope boresight so that the observed sky patch traces a large circle in the sky. The FWHM beamwidth at each of the observing frequencies is also given in Table 1. The receivers at the LFI frequencies are based on indium phosphide (InP) high electron mobility transistors (HEMTs) cooled to 20 K by a closed cycle hydrogen sorption cooler.  The HFI detectors include spider-web and polarization-sensitive bolometers operating at 0.1 K.  The LFI is sensitive to linear polarization in all channels while the HFI has 8 polarization sensitive detectors at each of its 4 lowest frequencies.

	The sensitivity in each of the Planck frequency bands is given in terms of the rms thermodynamic noise temperature in each corresponding beam area in Table \ref{tbl:originalspecs}. It is seen that the highest temperature sensitivities are at the lower frequencies. 

	The LFI instrument with its arrays of cooled pseudo-correlation radiometers represents an advance in the state of the art \citep{2000ApL&C..37..171B,2003astro.ph..7116M} over the sensitivity of COBE \citep{1990ApJ...360..685S} and WMAP \citep{2003ApJS..148...97B}. It is particularly designed to have maximum freedom from systematic errors both in total power and polarization. 

\begin{figure}
	\centering
	\includegraphics[angle=0,width=15 cm]{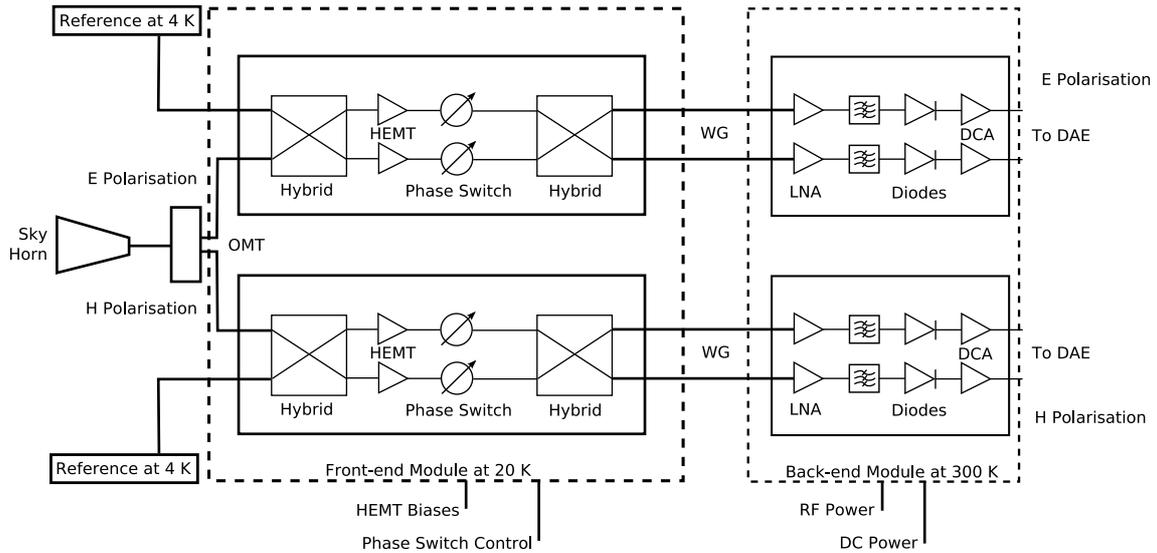} 
	\caption{The basic block diagram of the radiometer chain. The signals from the horn and reference load are fed to the hybrids via the ortho-mode transducer (OMT). The dashed line surrounding the hybrids, first amplifiers, and phase switches delineates the front end module (FEM). The dashed line surrounding the second amplifiers, filters, detectors and video amplifiers defines the back end module (BEM). The FEM and BEM are connected by four 1.5 m lengths of waveguide}
	\label{fig:blockdiagram}
	\end{figure}

	The overall rationale for the radiometer design, which differs  from the WMAP design \citep{2003ApJS..145..413J} is described in \cite{2009_LFI_cal_M2}. The LFI radiometer design is largely driven by the need to suppress the fraction of 1/f-type noise induced by gain and noise temperature fluctuations in the amplifiers which would be unacceptably high for a simple total power system. A differential pseudo-correlation scheme is adopted, in which signals from the sky and from a black-body reference load are combined by a hybrid coupler, amplified in two independent amplifier chains, and separated out by a second hybrid as shown in Figure \ref{fig:blockdiagram}.

	\begin{table*}
	\begin{minipage}[t]{\columnwidth}
	\caption[]{Original specification of the receivers and bolometers in Planck.}
	\label{tbl:originalspecs}
	\renewcommand{\footnoterule}{}  
	\begin{center}
	\begin{tabular}{ c c c c c c c c c c } \hline \hline
	Property & \multicolumn{9}{c}{Centre frequency (GHz)} \\
	& 30 & 44 & 70 & 100 & 143 & 217 & 353 & 545 & 857 \\ \hline
	LFI\footnote{LFI receives both linear polarizations and employs indium phosphide (InP) HEMT amplifers} or HFI\footnote{HFI units are spider-web and polarization--sensitive bolometers} & LFI & LFI & LFI & HFI & HFI & HFI & HFI & HFI & HFI \\
	Number of horns & 2 & 3 & 6 & 8 & 12 & 12 & 12 & 4 & 4 \\
	Linear polarization & Yes & Yes & Yes & Yes & Yes & Yes & Yes & No & No \\
	Resolution (arcmin) & 33 & 24 & 14 & 10 & 7.1 & 5.0 & 5.0 & 5.0 & 5.0 \\
	$\Delta T/T$ per pixel (Stokes $I$)\footnote{Goal in $\mu K/K$ 1$\sigma$ for square pixels of side given in the ``resolution'' row, achievable after 2 full sky surveys (14 months)}  & 2.0 & 2.7 & 4.7 & 2.5 & 2.2 & 4.8 & 14.7 & 147 & 6,700 \\ \hline
	\end{tabular}
	\end{center}
	\end{minipage}
	\end{table*}

	Each front end module (FEM) is located in the focal area of the Planck telescope where it is fed from two dual-profiled feed horns, one directed at the sky and the other at the 4-K reference load located on the surface of the high frequency instrument (HFI) unit. The FEM then receives the two input signals which pass through the first hybrid and are then amplified. After appropriate phase switching at $\sim$~8 kHz the signals pass through the second hybrid. The signals are then sent via $\sim$~1.5~m of waveguide to the back end module (BEM) at the 300~K station of the spacecraft where they are further amplified and then was measured using phase sensitive detectors. The BEMs at 30 and 44~GHz are described in detail by \citet[][this volume]{2009_LFI_cal_R9}, while the 70~GHz radiometers are described by \citet[][]{2009_LFI_cal_R10}. The output signal is the difference between the sky and the reference load with greatly reduced 1/f noise. This information is sent to the data acquisition electronics (DAE) module on the spacecraft.  

	The specification of the FEM parameters at 30 and 44~GHz is summarized in Table \ref{tbl:specifications}.

	\begin{table*}
	\begin{minipage}[t]{\columnwidth}
	\caption[]{Specification of the 30 and 44~GHz Front End Modules (FEMs) as required by the Planck mission.}
	\label{tbl:specifications}
	\renewcommand{\footnoterule}{}  
	\begin{center}
	\begin{tabular}{ l c c } \hline \hline
	~ & 30~GHz Channel (Ka band) & 44~GHz Channel (Q band) \\ \hline
	Centre frequency & 30~GHz & 44~GHz \\
	Number of FEMs & 2 & 3 \\
	Bandwidth  & $6.0$~GHz (20\% of 30~GHz) & $8.8$~GHz (20\% of 44~GHz) \\
	Noise temperature over band & 8.6 K, (6.1 K goal) & 14.1 K, (10.4 K goal) \\
	Gain (band average) & 30 dB min/33 dB max & 30 dB min/33 dB max \\
	Gain variation with frequency & $<$5 dB across band & $<$5 dB across band \\
	Radiometric isolation  & 10\%, (5\% goal) & 10\%, (5\% goal) \\
	FEM gain variation with ambient temperature & $<$0.05 dB/K & $<$0.05 dB/K \\
	FEM noise variation with ambient temperature & $<$0.8 K/K & $<$0.8 K/K  \\
	FEM gain variation with time & $<$1 dB /1000 hrs cold running & $<$1 dB /1000 hrs cold running \\
	1/f  knee frequency & $<$50 mHz, ($<$20 mHz goal) & $<$50 mHz, ($<$20 mHz goal) \\ \hline
	\end{tabular}
	\end{center}
	\end{minipage}
	\end{table*}

\section{FEM Design}
\label{design}

	\subsection{Physical Layout}

		The overall sensitivity of the Planck LFI, at each of its measurement frequencies, is largely determined by the design of its FEMs. In order to meet the requirements of the Planck project it was necessary to achieve amplifier noise temperatures lower than previously achieved with multi-stage transistor amplifiers. As the instrument is designed to measure the sky temperature in orthogonal linear polarisations, the input ports of the FEM connect to the instrument's feed horn via an ortho-mode transducer (OMT). Referring to the block diagram in Figure \ref{fig:blockdiagram}, each FEM comprises two input hybrid couplers (one per polarisation), four low noise amplifiers (LNAs), four phase switches, and two output hybrids.

		\begin{figure}
		\centering
		\includegraphics[angle=0,width=15 cm]{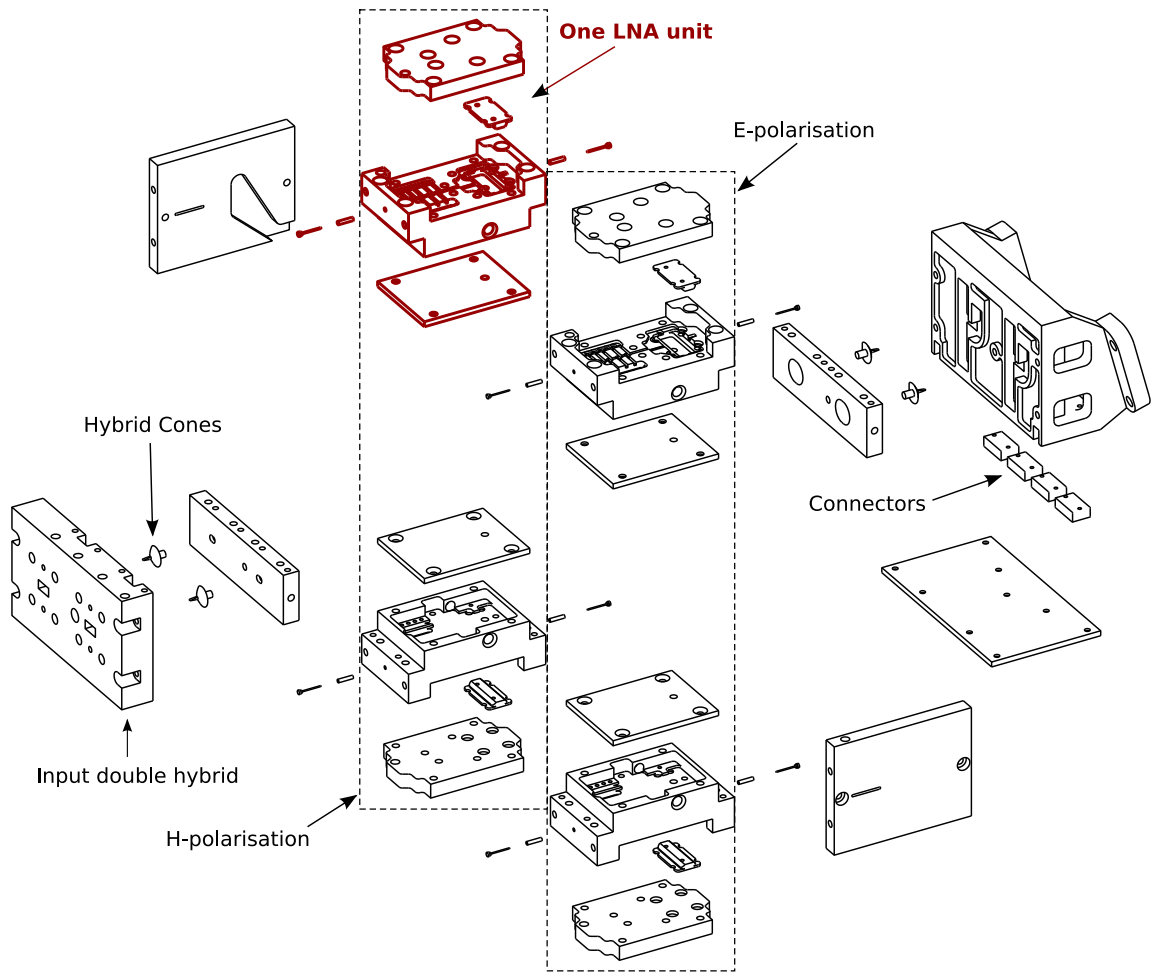} 
		\caption{Exploded diagram of FEM showing input and output double hybrids, 4 low noise amplifiers grouped one pair to each polarisation channel and associated side plates and divider plates.  A single LNA is outlined in red, and the polarization pairs are outlined in dashed lines.}
		\label{fig:exploded}
		\end{figure}

		To achieve the lowest possible noise temperature, it was necessary to minimise any loss in front of the LNAs and this, together with other performance criteria, pointed to the use of a waveguide hybrid coupler at the input. The specification of the waveguides to connect the FEM to the BEM dictated the use of an essentially identical coupler for the output. It was decided at an early stage that in order to fit all the above elements of the FEM into the restricted volume available around the HFI instrument, a highly integrated design would be required. This was not readily commensurate with the mix of transmission techniques desired, or with the desirability of being able to test each item individually. A multi-split-block solution was adopted, where the four LNA/phase switch modules were mounted to end plates, side by side pairs of LNAs being arranged above and below, in a mirror-image format around the centreline. The end plates formed one waveguide face of each hybrid coupler, to which a block containing the machined waveguides was attached. The only difference in the design of input and output hybrids was in the routeing of the H-plane waveguides to the external interfaces. Figure \ref{fig:exploded} shows an exploded view of the FEM assembly.

		Mass was a major consideration. Gold-plated aluminium alloy (6082-T6 grade) was used, but part thicknesses, mounting screw positions and torque levels had all to be considered to avoid distortion of the critical interfaces, where imperfections could result in unwanted signal leakage, crosstalk and loss. To reduce mass, unnecessary metal was machined away, leaving good contact around waveguide joints, with sufficient stiffness especially near mounting points.

		The microwave absorbing surfaces of the 4~K matched loads are mounted around the periphery of the HFI instrument. To access the two load horn waveguides conveniently, the H-plane hybrid outputs terminate on the inner surface of the FEM as assembled in the LFI. The FEM mounting flanges were machined at a precise beam angle for the sky horns.

		With up to 20 low noise transistors and 8 phase switch diodes per FEM, all operating at 20~K, several requirements, some conflicting, had to be considered in the design of the bias circuitry.  Since the expected in-flight cooling power available to the entire Planck focal plane unit is only $\sim$1.1~W, all power dissipation had to be kept to the absolute minimum. This set the requirement to use InP active devices operated in power starved mode.  When testing the amplifiers it is necessary to control each individual gate and the drain voltages. When the best values were found potentiometers were used to common most of the control wires to minimise the number of supply lines. Since the FEMs are driven by $\sim$~1.5~m of wire length, it was necessary to terminate each wire in the FEM with electrical protection devices. Nuclear hardened light-emitting diodes (LEDs) were found to be suitable and can remove any sharp spikes which are induced on to these long lines.

		The maximum allowable dissipation per FEM was dictated by the cooling power of the Planck 20~K sorption cooler \citep{2009_LFI_cal_T2} and the thermal losses down the FEM/BEM connecting waveguides and wiring. Such critical limitations were met with drain voltages as low as 0.6~V. A dedicated bias test power supply with a high level of protection, capable of supplying up to 5 drain, 5 gate and 2 phase switch voltages to each of four LNAs was designed and built. Both gate and phase switch supplies were adjustable for either polarity.

		Assembly and testing (outside the cryostat) was done under clean room conditions using pre-cleaned parts.  In order to avoid any leakage between the radiometers, it was necessary to assemble the different parts of the modules with great care as to alignment of all faces and flanges.

	\subsection{ Hybrids}

\begin{figure}
		\centering
		\includegraphics[angle=0,width=15 cm]{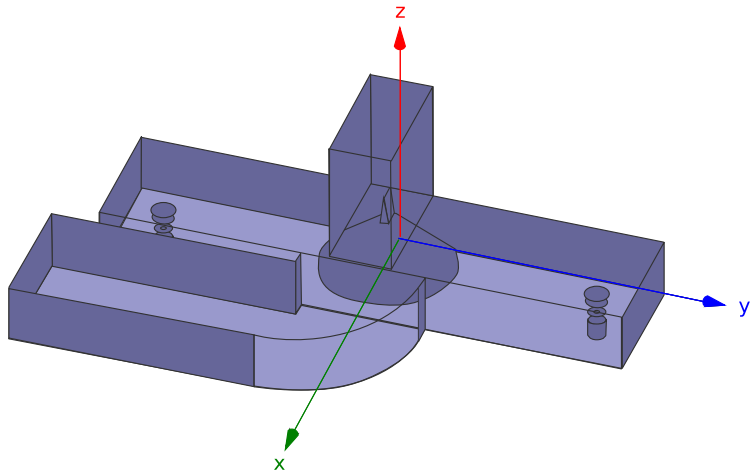}
		\caption{Magic T Hybrid waveguide as realized in Planck FEMs}
		\label{fig:magicT}
		\end{figure}

\begin{figure}
		\centering
		\includegraphics[angle=0,width=15 cm]{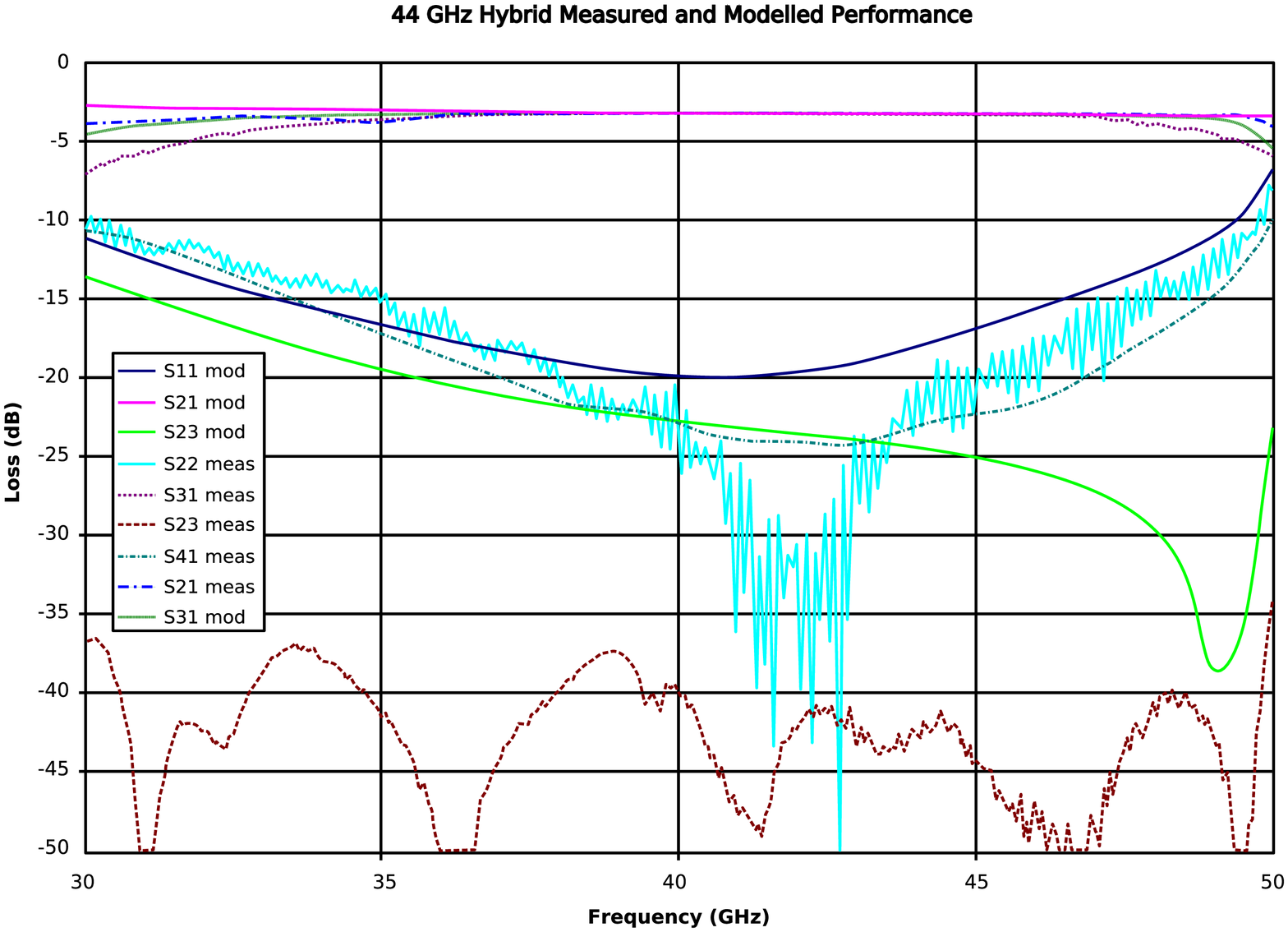}
		\caption{44~GHz measured hybrid S-parameters. All unwanted cross-talk parameters are less than 15~dB. The wanted signals eg S21 are imperceptibly different from the expected 3~dB value.}
		\label{fig:sparameters}
		\end{figure}

		The input and output hybrid couplers utilise the conventional ``magic tee'' configuration and differ only in the way that the input and output waveguides are configured in order to interface with other parts of the radiometer (Figure \ref{fig:magicT}). At each junction is a matching probe centred between the balanced arms, but offset away from the H-plane waveguide interface. Ansoft's High Frequency Structure Simulator (HFSS) electromagnetic modelling software was used to determine the dimensions and offset of the probe for optimum hybrid performance. In all cases, the performances were checked with model jigs. The hybrids, being an integral part of the FEM body, were realised using split-block construction, machined from aluminium alloy then gold-plated. In operation the FEM is cooled to 20~K. The combined effect of the waveguide design, and the low temperature and consequent high conductivity of the gold, means that the input hybrids contribute negligible noise to the system. A typical plot measured at room temperature is shown in Figure \ref{fig:sparameters}. The insertion loss above the nominal 3~dB split was less than 0.1~dB.

		The balanced arms of the hybrid are terminated in waveguide-to-coaxial transitions where the short-circuited end of each transition forms part of the hybrid, while the probe forms part of the LNA. This is shown in the FEM exploded view, Figure \ref{fig:exploded}. As with the hybrids, performance of the transitions was optimised using both HFSS modelling and test model measurements. The amplifiers were tested with jigs to convert from stripline to waveguide. These consist of a short piece of waveguide with a back short into which the LNA pins sit. The loss of the jig system compared to the hybrid connection when the LNAs were in the FEM contributed less than 1~K, which was our measurement error. Thus the extra loss of the hybrids must be less than 0.2~dB. This result was extremely important for the performance of the whole FEM as the first hybrid comes before the LNAs.

	\subsection{ Low Noise Amplifiers}

		\subsubsection{ LNA design}

\begin{figure}
			\centering
			\includegraphics[angle=0,width=15 cm]{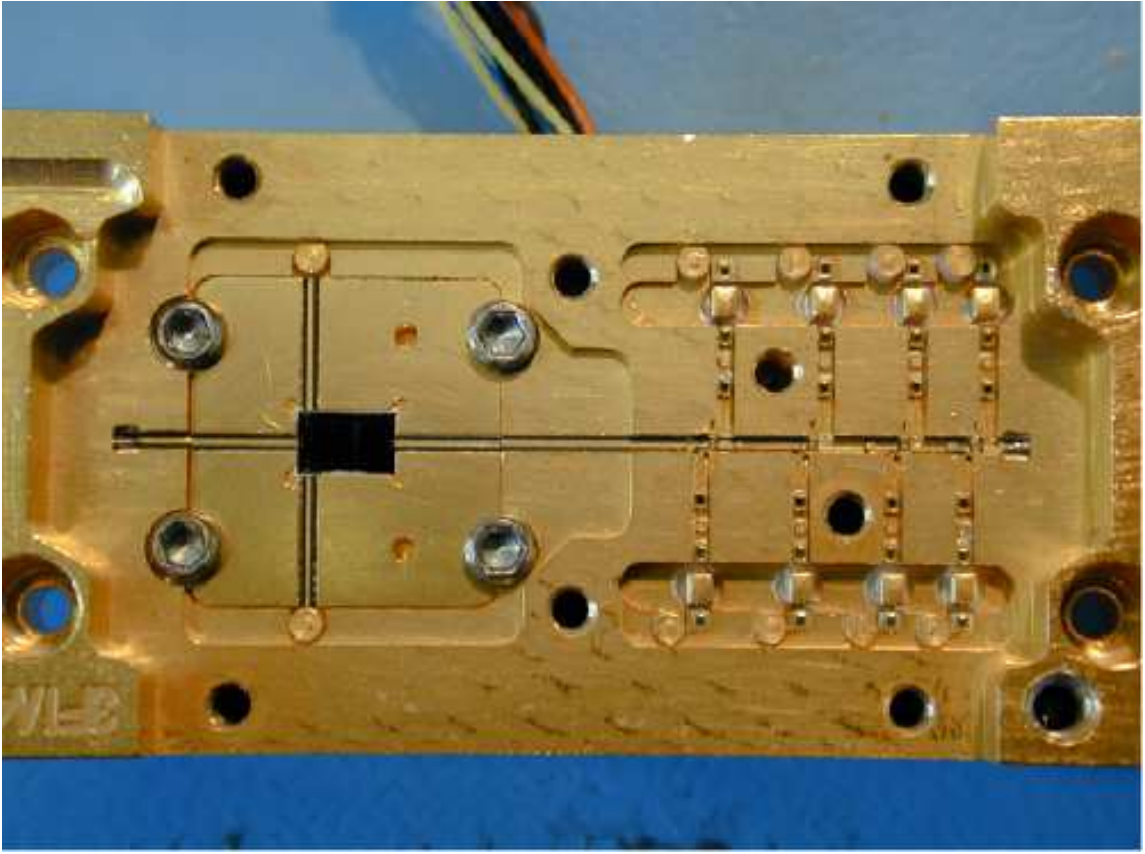}
			\caption{30~GHz FM LNA and phase switch. The four transistors can barely be seen in the RF channel running from from right to left. The phase switch is seen as a black rectangle on the left. The bias voltages to the transistors and control voltages to the phase switches can seen from top to bottom of the figure.}
			\label{fig:fmLNA}
			\end{figure}

			Each of the four outputs from the input hybrids couples via a tuned probe and short length of coaxial line to an LNA constructed using microwave integrated circuit (MIC) techniques. Each LNA output connects to the output hybrid via a phase-switch. Figure \ref{fig:fmLNA} shows the 30~GHz LNA and phase switch in the FEM RF section.

			The designs of the Ka-band and Q-band HEMT-based amplifiers drew significantly on the work of  \citet{2000Pospieszalski}. A 30~GHz NRAO design had been further developed over a number of years for use at cryogenic temperatures in the Jodrell Bank Observatory (JBO) Very Small Array (VSA) receivers operating on Tenerife. This amplifier was used as the starting point for the Planck 30-GHz LNA. The cryogenic hardening was critical for the Planck amplifiers, as was the ability to survive the vibration due to the launch. Due to the smallness and lightness of the RF components, g-forces are negligible at the frequencies generated during the Ariane launch, and this was not in practice an issue (see Section 4.2).

		\subsubsection{HEMT and MIC design}

			Indium phosphide HEMTs manufactured by TRW (now NGST) are used as the active elements, as they provide extremely low noise performance with minimal power dissipation (as low as $5$~mW per LNA at 30~GHz), which is vital for this space-borne project with limited cooling capacity. Each LNA provides a minimum of $30$~dB of gain, and the pairs of LNAs in each channel are matched in both gain and phase over the 20\% bandwidth. In view of the need for the lowest achievable noise temperature, the final 30-GHz design uses 80~$\mu$m gate width, 0.1~$\mu$m gate length InP HEMTs provided to JBO by JPL Pasadena from the Cryogenic HEMT Optimisation Program (CHOP) CRYO3 wafer run for the first stage and other 80~$\mu$m gate width devices for the remaining stages. The 44-GHz design uses 60~$\mu$m gate width HEMTs from the same CRYO3 wafer for the first stage and 80~$\mu$m CRYO3 for the remaining stages.

			The CRYO3 wafers were found to be absolutely critical for this project. Not only did they give the lowest noise, but also the particular wafer had an unusually thin passivation layer which proved essential for the highest frequencies.  At the time of the Planck construction this particular CRYO3 wafer was the only one in existence with these particular properties.

		\subsubsection{LNA construction}

\begin{figure}
			\centering
			\includegraphics[angle=0,width=15 cm]{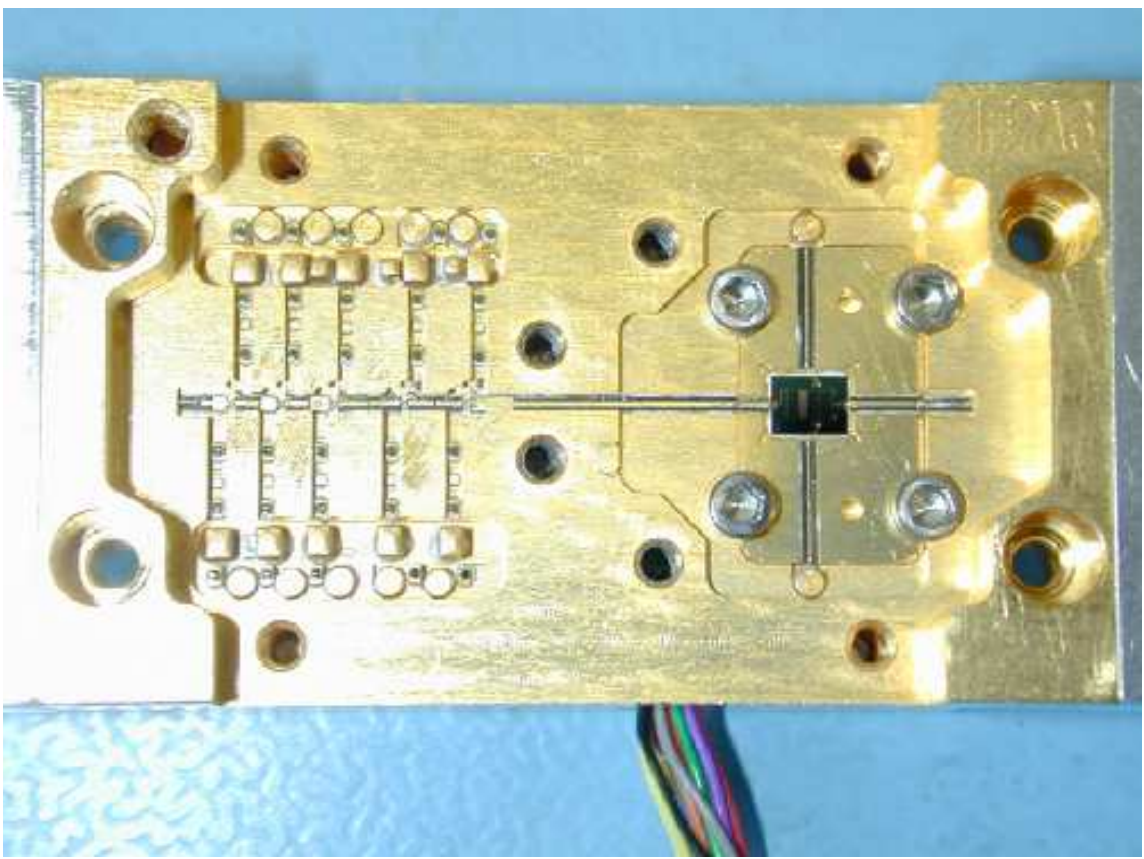}
			\caption{44~GHz (Q-band) LNA, RF side.}
			\label{fig:LNARF}
			\end{figure}

			Figure \ref{fig:LNARF} shows the upper (RF) side of a Q-band (44~GHz) LNA/phase switch assembly, where the probes forming a part of each transition to the symmetrical waveguide arms of the hybrid can be seen projecting from each end. The RF input is on the left-hand side, with the phase switch mounted on a carrier towards the right-hand side of the chassis. The remainder of the RF side of the LNA, including the integral phase switch, is realised in MIC microstrip format using a pure PTFE substrate (CuFlon) for lowest loss. 

			The EM fields are contained as far as possible within a narrow channel to minimise the production of higher-order waveguide modes. The RF circuit boards are just visible in Figure \ref{fig:LNARF} as thin grey lines with gold tracks down their centres. Effective grounding of these boards was found to be very important and silver conductive glue was used in the area beneath each board. 

			Some components of the bias circuits are sited on the RF side. To minimise the complexity of supply wiring, each amplifier was supplied with a single drain, two gate and two phase switch voltages. One of the gate inputs supplied the critical first stage while the other input was common to the remaining stages. To ensure that each HEMT received the correct gate bias as determined by the cold tests, a potentiometer, made up of resistor array chips, was used to drop the common voltage to that required. These chips were by-passed throughout the testing phase, and only bonded into circuit after the final cold test described above. 

			The remainder of the DC section, comprising potentiometer networks for the gate supplies, a distribution PCB, and protective LED diodes, is mounted in a compartment below the RF section, with feed-through pins supplying power to the ladder networks above. From the DC interface which is a Nanonics connector, fine PTFE-covered wires feed supply voltages to the distribution PCB. These are screened with braid in the 30~GHz FEM and shielded by a cover plate in the 44~GHz FEM.

\begin{figure}
			\centering
			\includegraphics[angle=0,width=15 cm]{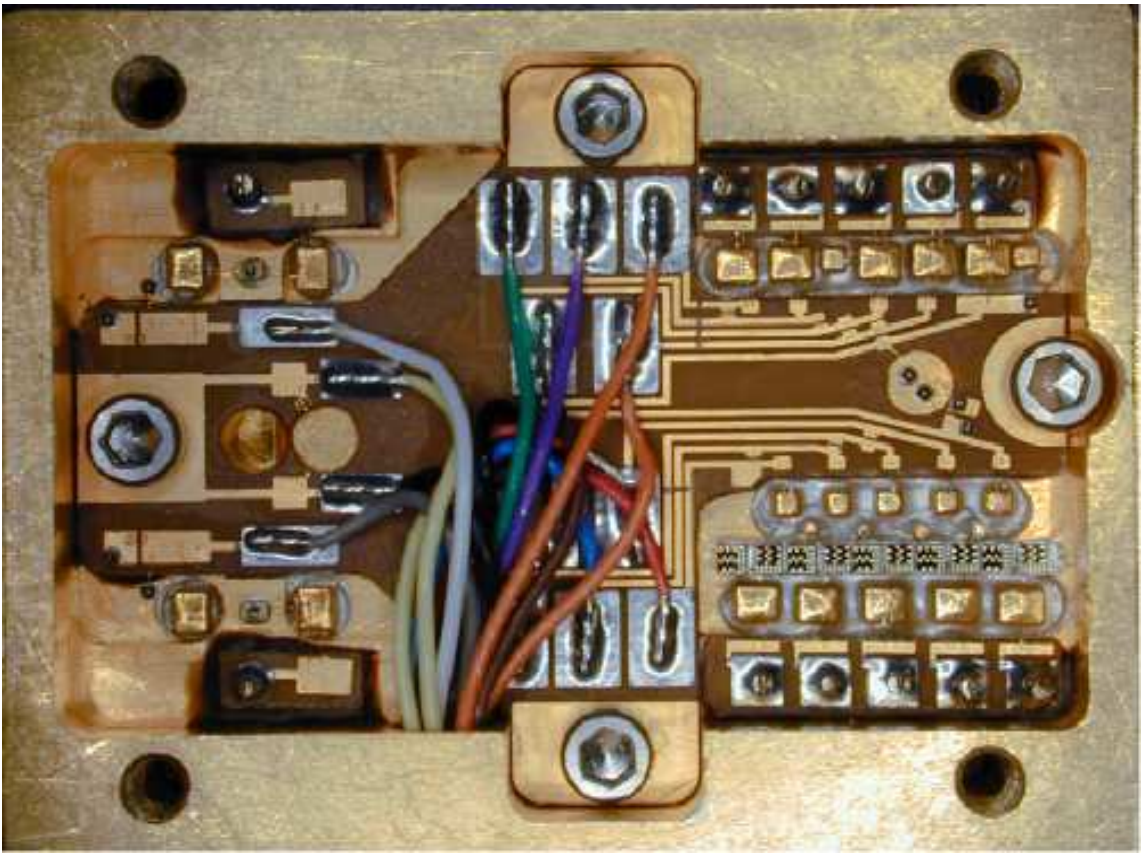}
			\caption{Q-band LNA DC side}
			\label{fig:LNADC}
			\end{figure}

			Two sets of amplifiers were built for each FEM, one set wired from the opposite side to the other. To the right side of Figure \ref{fig:LNADC}, the LNA bias circuitry can be seen, and to the left, the phase switch bias.

			During transit and handling, protective covers were fitted at all times to prevent static or mechanical damage.

		\subsubsection{ LNA Test set-up}

			A square cross-section test cryostat with easily removable outer and inner lids was built for LNA testing. Apertures were machined around the periphery to allow a wide range of input, output or bias interfaces to be fitted using custom-built interface plates. For the standard noise/gain measurement, a circular horn was attached to the LNA input behind and close to a MylarTM (PET) window. This allowed the hot and cold load technique to be applied successively outside the cryostat. The physical temperature of the test cryostat was typically 18~K and stable to 0.5~K.

			The output was measured using a double sideband mixer, an HP8970B noise figure meter and an HP8350B sweep oscillator. 

\begin{figure}
			\centering
			\includegraphics[angle=0,width=15 cm]{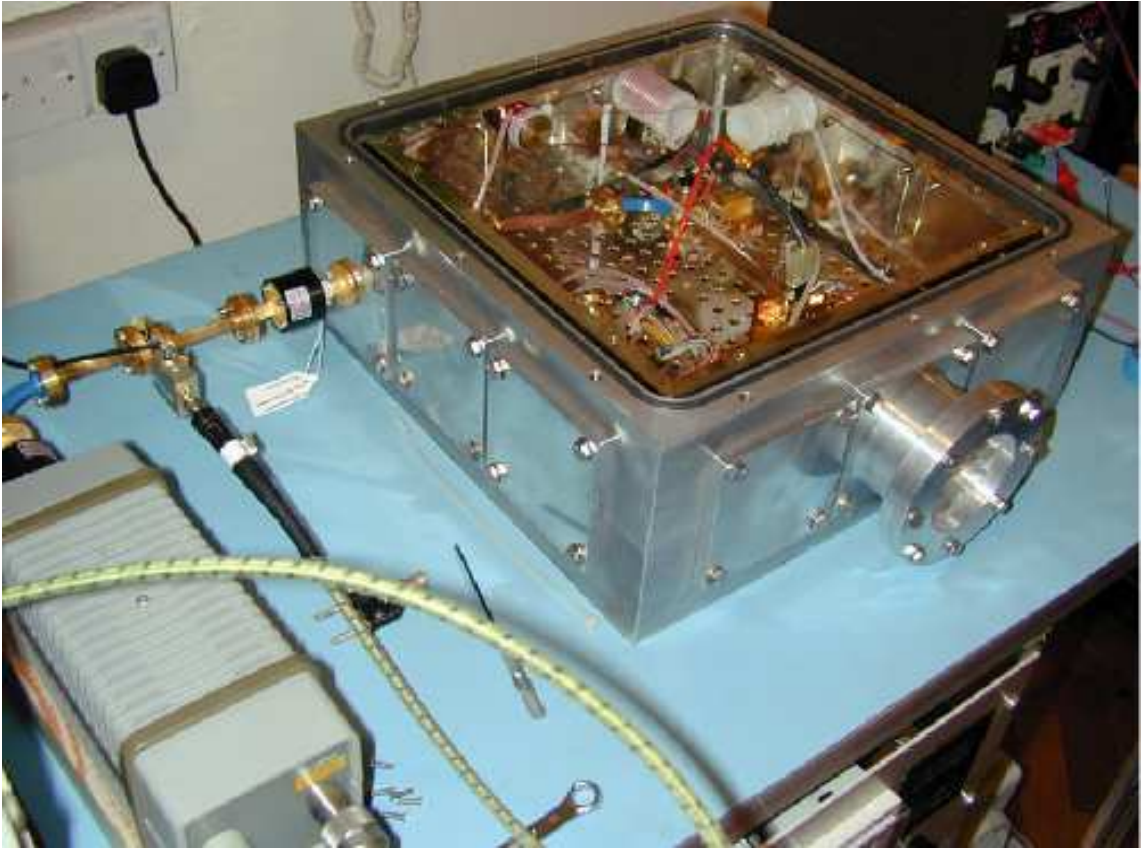}
			\caption{LNA test cryostat, configured for Q-band LNA measurements}
			\label{fig:LNAcryostat}
			\end{figure}

			The test set is shown in Figure \ref{fig:LNAcryostat}. The test computer stepped the test frequency of the equipment across the band for each temperature, reading the noise power at each step.

			The gain and noise performance were measured using reference loads comprising copper cones lined with microwave absorber, one at room temperature, the other at liquid nitrogen temperature. Each was placed in turn in front of the mylar window covering the cold horn inside the cryostat.  The results were corrected for the extremely small estimated losses in the horn, transition and window.  Any devices with measurable gate leakage currents were detected and removed at this point

			In an attempt to standardize noise temperature measurements, a 44~GHz LNA built at NRAO, in the same way as the WMAP amplifies were built, was compared with the Planck amplifiers. This LNA was flown to JBO and tested on the above test system, and a JBO LNA was flown to NRAO and tested on their test set-up. The conclusion was that the systems were in agreement to $1$~K noise temperature.

		\subsubsection{Modelled and actual performances of LNAs}

			The design software used for the LNAs was MMICAD V2. The starting point for the HEMT was the ``FETN'' model \citep{1989ITMTT..37.1340P} in the element list, with small modifications at times to suit the device data available.

			The remainder of the circuit was modelled with microstrip elements, and lumped elements where dimensions were considered sufficiently short. The circuits were optimised for gain, bandwidth, gain variation, stability and input/output return losses. Phase matching of LNAs was effected after completion of individual LNAs, by small experimental changes to biases and bond lengths. Although a recheck was necessary, this normally had little effect on the other parameters. 

			During development, device data were also obtained  by room temperature (RT) Vector Network Analyser (VNA) measurements up to 110~GHz using on-wafer probing.

			A prototype LNA, with well-matched WR28 transitions on input and output, gave excellent noise results. However, when the same LNA design was utilised in the 30~GHz front end module, with WR28 hybrid couplers on input and output, the noise performance was significantly inferior to that of the prototype LNA. Loss of the gold-plated waveguide hybrid was calculated to be negligible, so the source of the additional noise had to be sought elsewhere. The impedance match `seen' at the input of the LNA was thought to be a possible cause of the problem. Whereas the input and output of the prototype LNA were well-matched, i.e. the LNA `saw' an impedance very close to 50 ohms, this was not the case for the hybrid coupler. This effect is described by the well known relationship \citep{1960PIRE..48..66}:

			\begin{equation}
			T_\mathrm{e} = T_\mathrm{min} +T_{0} \frac{R_\mathrm{n} }{G_\mathrm{s} } \left|Y_\mathrm{s} - Y_\mathrm{min} \right|^{2}.
			\label{eqn:Te}
			\end{equation}

			$T_\mathrm{e}$ is the effective noise temperature of the LNA

			$T_\mathrm{min }$ is the minimum noise temperature of the LNA (given optimum impedance matching)

			$T_{0}$ is the standard reference temperature (290 K).

			$R_\mathrm{n}$ is the equivalent noise resistance of the LNA.

			$G_\mathrm{s}$ is the source conductance.

			$Y_\mathrm{s}$ is the source admittance as `seen' by the LNA input.

			$Y_\mathrm{min}$ is the optimum source admittance.

			As the LNA connects to the waveguide via a built-in transition (microstrip-coax-waveguide) it is not possible to directly measure the impedance at the interface. Instead the hybrid coupler (Figure \ref{fig:magicT}), including transitions, was modelled in HFSS to produce a 4-port S parameter matrix.

			The 4-port scattering matrix for the hybrid coupler was  used in a circuit simulator (Agilent ADS) to model the performance of the first stage of the 30~GHz LNA. It was then possible to optimise the length of the first stage gate bond for minimum noise temperature, i.e. making $Y_{s}$ close to $Y_{\min }$. Good agreement was finally obtained between the model and the measured performance of the amplifier.

\begin{figure}
			\centering
			\includegraphics[angle=0,width=15 cm]{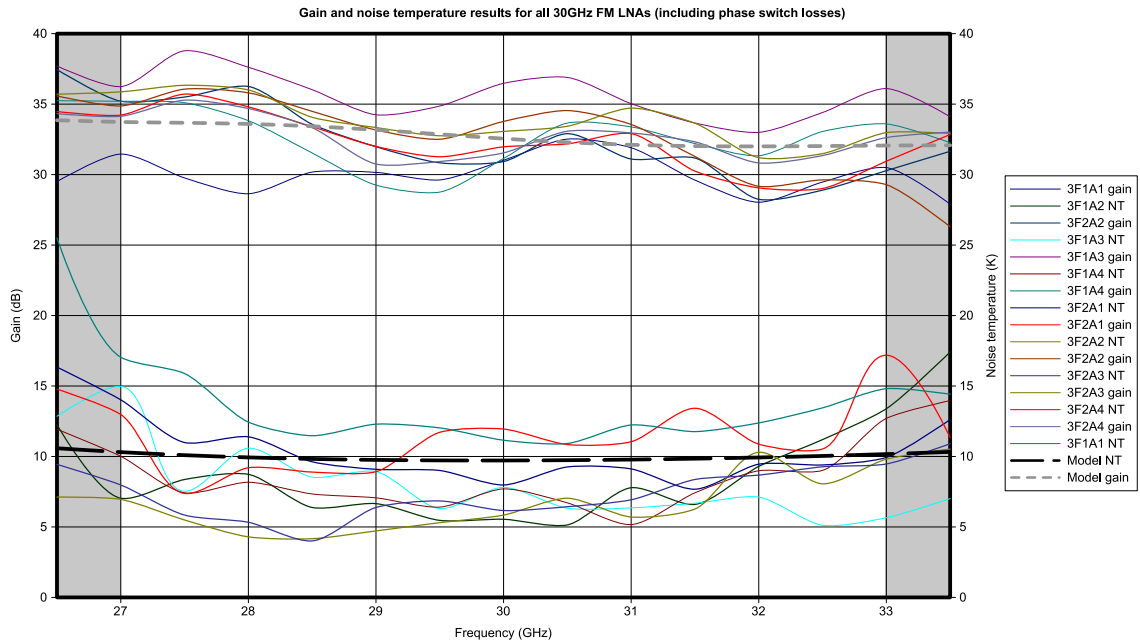}
			\caption{Gain and noise temperature performance for all 30~GHz FM LNAs (including phase switch losses). The modelled parameters are indicated by a bold dashed line. The Planck band is indicated by the white central part of the plot.}
			\label{fig:gain30}
			\end{figure}

\begin{figure}
			\centering
			\includegraphics[angle=0,width=15 cm]{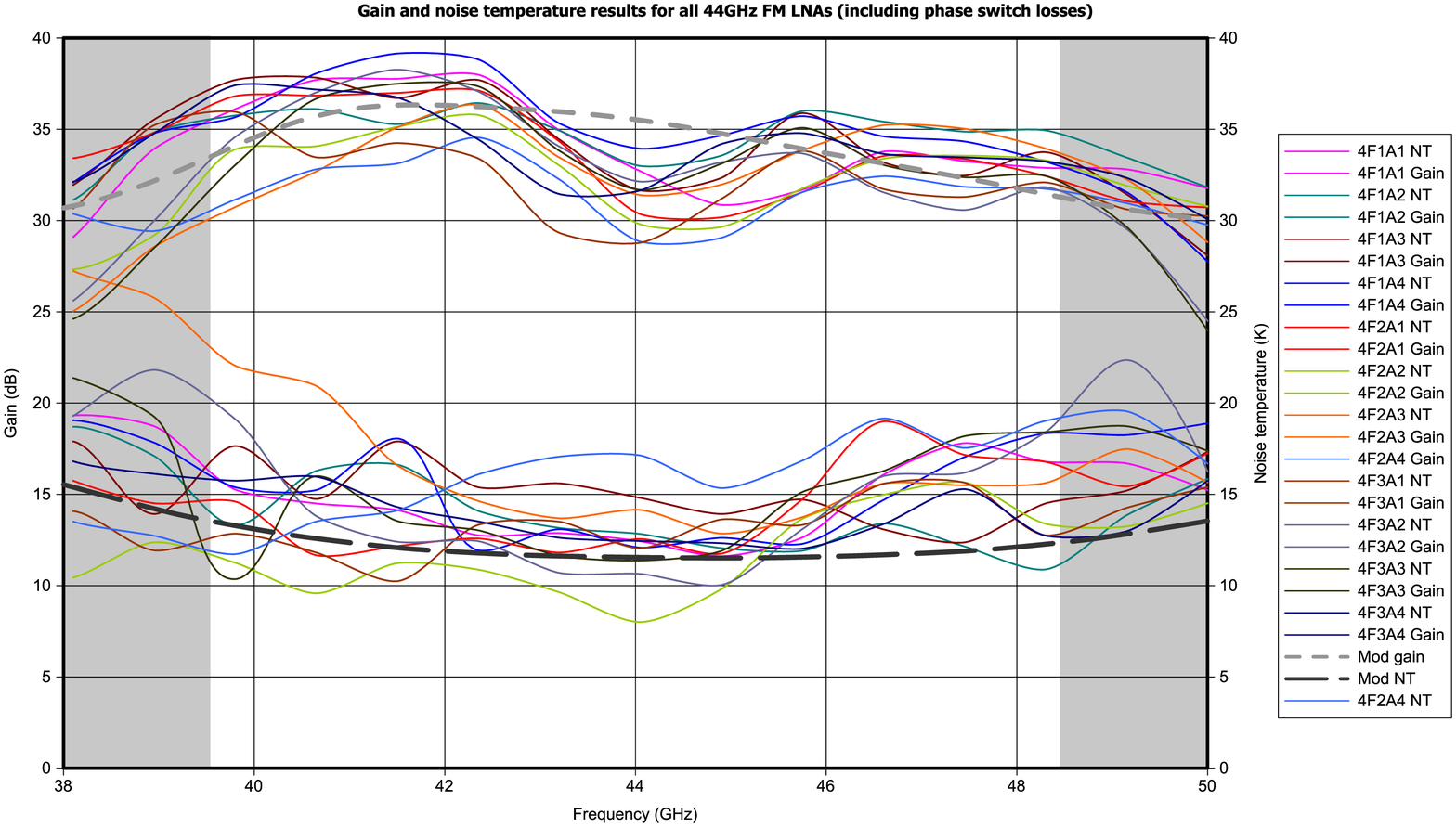}
			\caption{Gain and noise temperature performance for all 44~GHz FM LNAs (including phase switch losses). The modelled parameters are indicated by the bold dashed lines. The Planck band is indicated by the central white part of the plot.}
			\label{fig:gain44}
			\end{figure}

			Modelled stability has its limitations, as the HEMTs have gain capability well over $100$~GHz where model accuracy is poor. It was found necessary to use lossy, low-Q resonators in each drain circuit. These resonators comprised a bond wire to a resistive film deposited on a quartz substrate, the inductive bond being tuned in length to damp out instabilities, often close to $100$~GHz. At 44~GHz, it was also found advantageous to use the same technique at a frequency below the LNA band to flatten the gain response. Gain and noise temperatures with the latest modelled results are shown in Figures \ref{fig:gain30} and \ref{fig:gain44}. These modelled results and measurements agree. Essentially the gains is optimised across the correct Planck bands, the noise temperatures are a minimum in these bandwidths and at both frequencies the amplifiers achieve the requirements for the LNAs as indicated in the scientific requirements (see Table \ref{tbl:originalspecs}).

	\subsection{ InP MMIC Phase switches }

		\begin{figure}
			\centering
		\includegraphics[angle=0,width=15 cm]{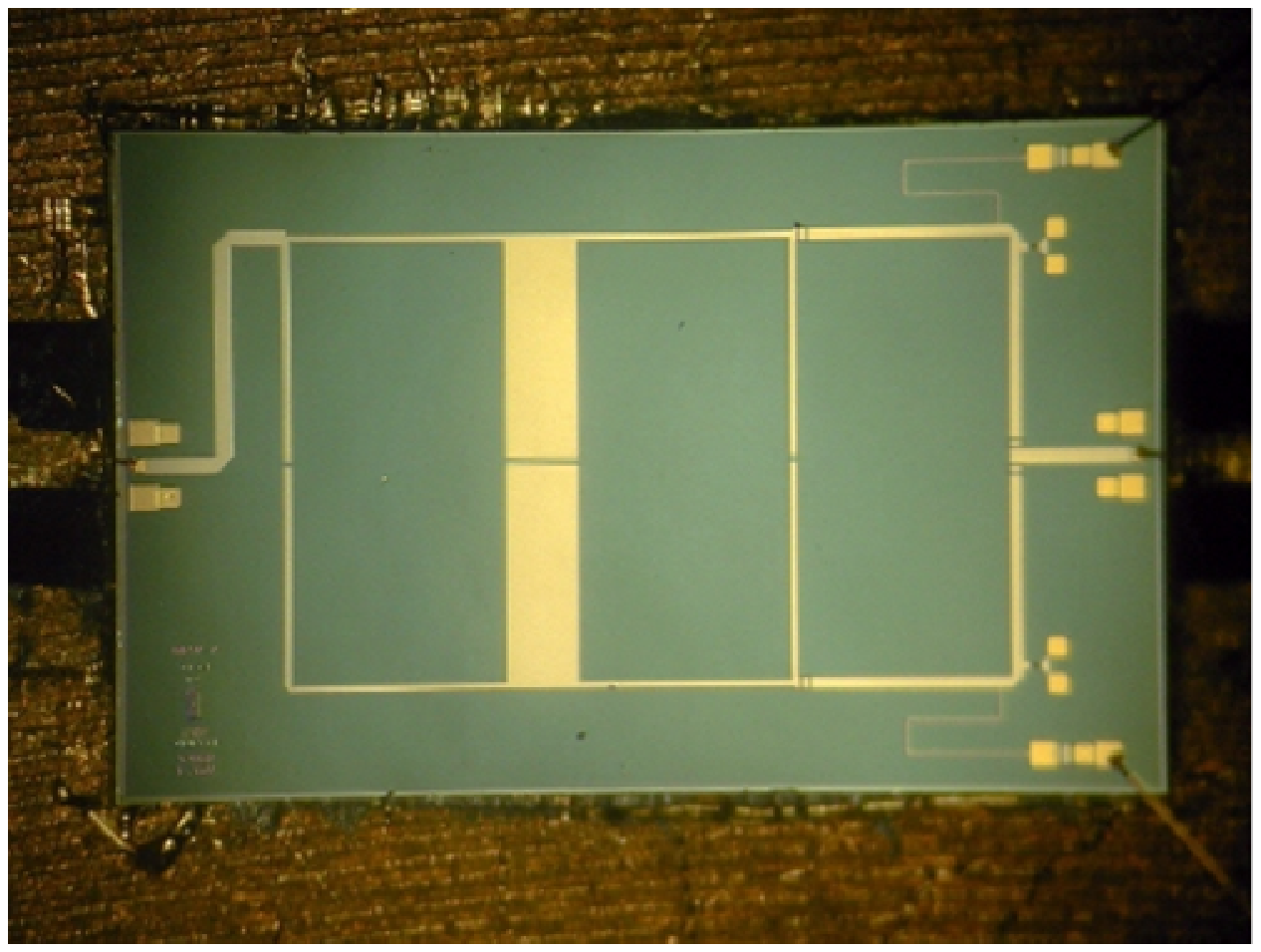}
		\caption{Photograph of the InP MMIC Phase switch manufactured at NGST, USA.}
		\label{fig:MMIC}
		\end{figure}

		Figure \ref{fig:MMIC} shows the 30~GHz version of the InP phase switches which are integrated in the PLANCK LFI FEMs at all frequencies. The basic design \citep{2003Hoyland} using a double hybrid ring configuration was scaled to the three centre frequencies $30$, $44$ and $70$~GHz. Phase control is obtained through two shunt PIN diodes attached to the circuit through quarter-wave length striplines. The performance of each band has been presented previously \citep{2003Hoyland}. The phase and amplitude match are plotted on a separate plot in Figure 12. The excellent amplitude ($<$0.05~dB) and phase match (180\degr$\pm1$\degr) can be seen across the entire 27-33~GHz band. Return loss is less than -10~dB over the band and the insertion loss is better than -2.5~dB for a non-starved mode of operation of 500~$\mu$W power consumption.

		Apart from the consistently high microwave performance of the phase switch  it proved critical that the phase switch was capable of operation with $<$ 1~mW power consumption at cryogenic temperatures. This was achieved by using the NGST InP PIN diode process which gives state-of-the-art performance. 

		\begin{figure}
		\centering
		\includegraphics[angle=0,width=15 cm]{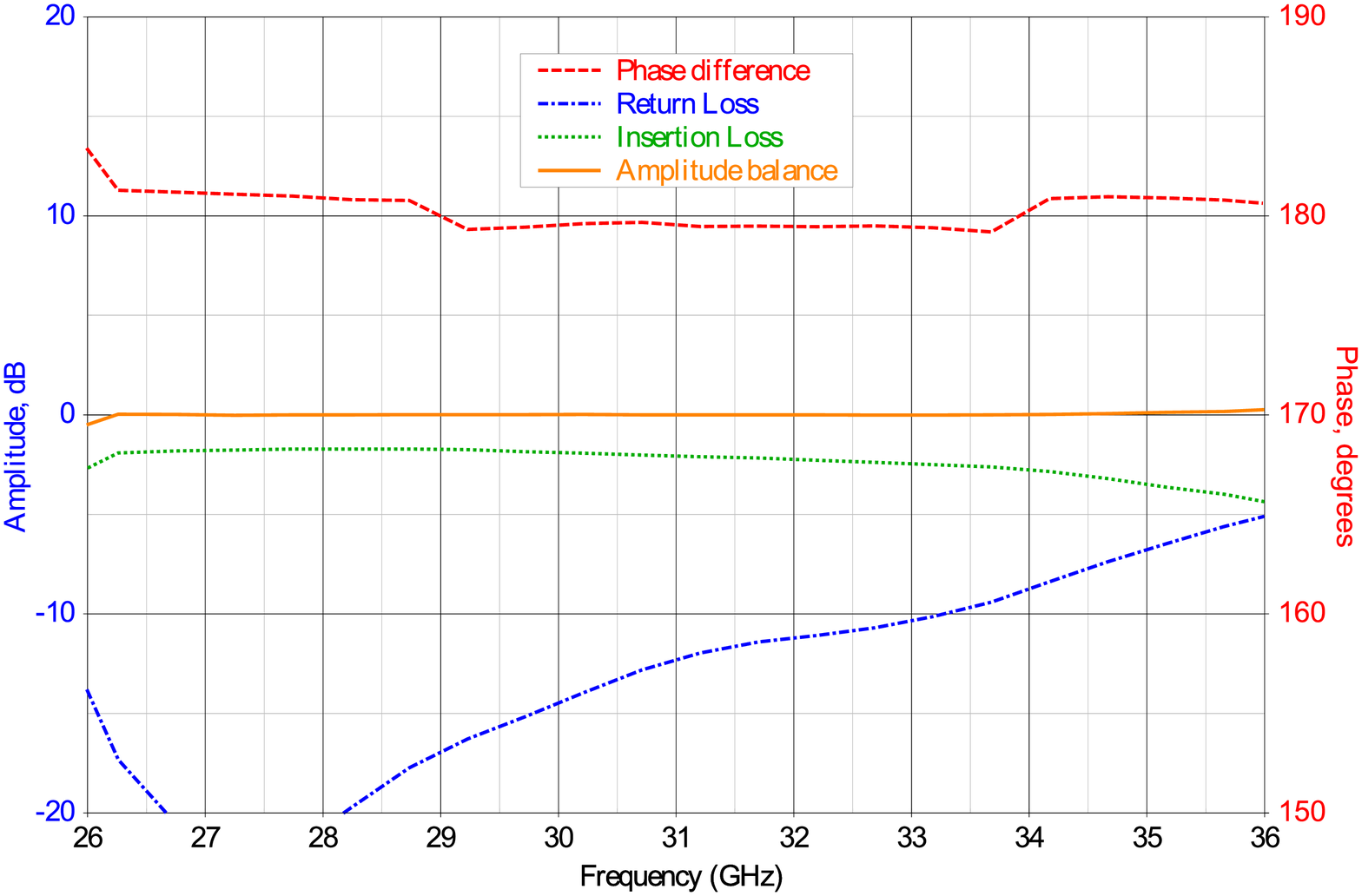}
		\caption{Graph to show typical phase match, return loss, insertion loss and amplitude balance of the 30~GHz InP phase switch}
		\label{fig:match}
		\end{figure}

		Although the phase switch is integral with the LNA in the final configuration, it is mounted on a small plated metal carrier bolted within the RF section of the LNA and then bonded across the minimal gaps. This arrangement permits the removal and separate jig testing of a phase switch or, by fitting a 50~ohm line to a dummy carrier, the separate testing of an LNA module. For rigidity, the carriers are of substantial thickness except at the interfaces, where they are stepped to minimise RF discontinuities.

		A MMIC version of this circuit, using MMIC PIN diodes manufactured by NGST (NRC) was subsequently used in all qualification model (QM) and flight model (FM) amplifiers. An early problem with attachment of these large chips was resolved and the excellent performance could be maintained down to below 0.5~mA current.

	\subsection{LNA optimisation}

		\subsubsection{Gain and noise temperature}

			\begin{figure}
			\centering
			\includegraphics[angle=0,width=15 cm]{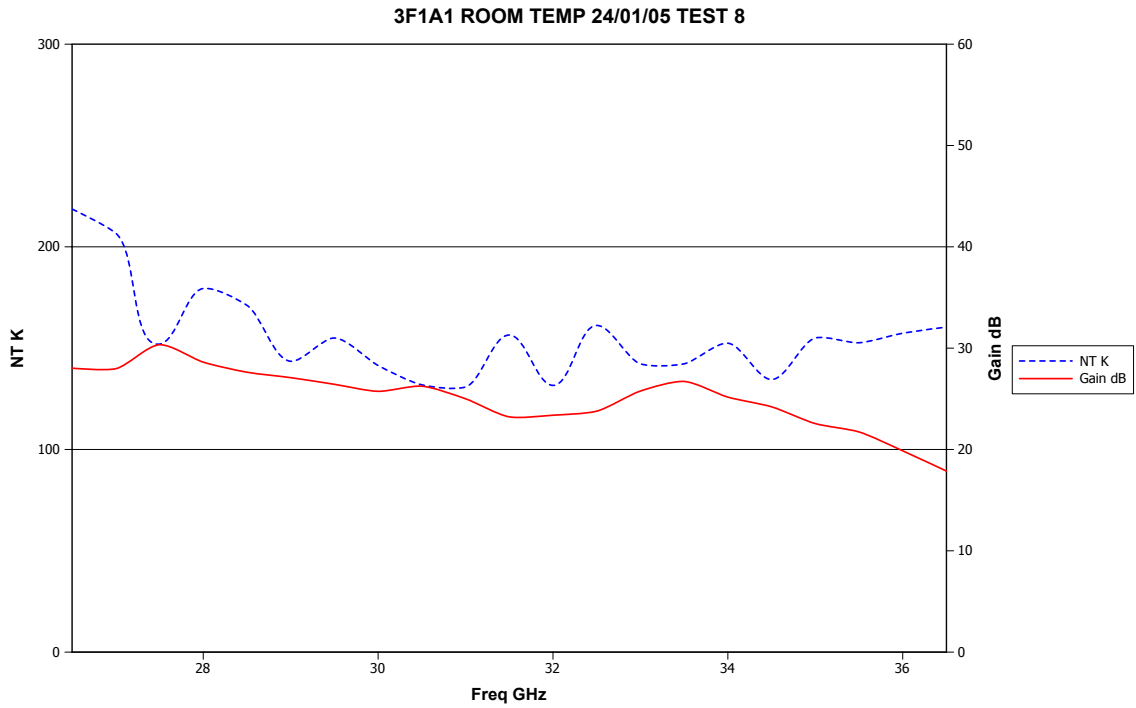}
			\caption{3F1A1 room temperature gain and noise temperature across the Planck band (27-33~GHz), CRY03 in 1st stage with 15~thou bonds.  The gain is lower and the noise temperature much higher than in the corresponding 20~K measurements.}
			\label{fig:3F1A1}
			\end{figure}

			With the assembled amplifier jig-mounted, S-parameters were initially measured on the VNA using common drain voltages and currents for all amplifiers, and in both phase switch states. These results gave room temperature (RT) gain, phase, input and output return losses, and isolation. 

			At this stage, the gain profile was optimised by altering the biases, the last two or three stages being tuned for gain performance (4 stages in total at Ka-band, 5 at  Q band), and the first two stages for best noise performance. Once the results were acceptable, the LNA was mounted in the test cryostat at room temperature and, starting from the same standard biases, tuned for good noise performance and gain across the band (see, for example, results for amplifier 3F1A1 in Figure \ref{fig:3F1A1}). This was usually achieved by modification of gate or source bond lengths. The gain responses of potential pairs of amplifiers were matched as closely as possible. 

\begin{figure}
			\centering
			\includegraphics[angle=0,width=15 cm]{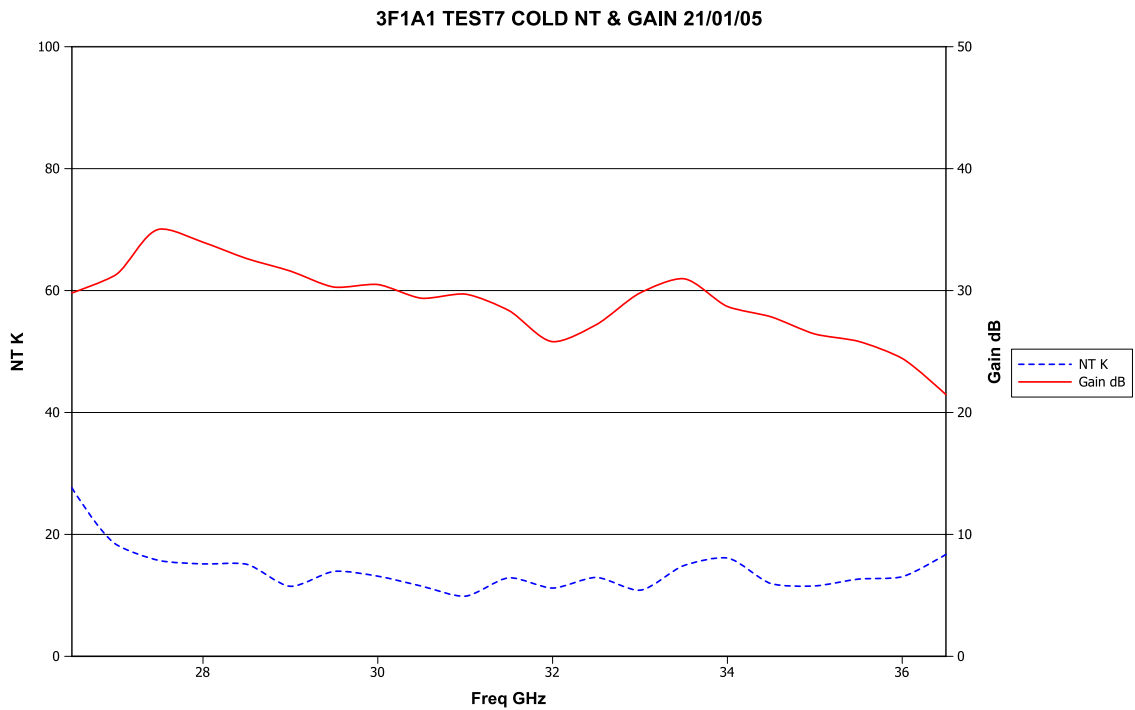}
			\caption{3F1A1 gain and noise temperature, measured at 20~K, CRYO3 in 1st stage}
			\label{fig:3F1A1NT}
			\end{figure}

			Once the room temperature tuning was deemed satisfactory, the amplifier was cooled to approximately 20~K, the biases reset to common, but lower, drain voltages and to a set of known initial drain currents which typically resulted in good cryogenic performance. Stability was confirmed over a range of bias conditions. Small adjustments to first and second stage gate voltages were used to finally optimise noise performance (see results for the same amplifier, 3F1A1, at 20~K, Figure \ref{fig:3F1A1NT}). Any changes other than bias adjustments were extremely time-consuming at this stage, as they necessitated additional cryostat heating and cooling cycles. Most of the amplifiers needed several room temperature and cryogenic test iterations during the tuning process.

		\subsubsection{ Amplitude and Phase Matching}

			Once amplifiers had been paired up for use in each channel of a FEM, the room temperature phase responses were compared, and modifications made as needed to equalise the absolute mean phase across the band.  The LNAs in each radiometer need to be well matched in both amplitude and phase for best radiometer performance, isolation etc.

\begin{figure*}
			\includegraphics[angle=0,width=17cm,clip]{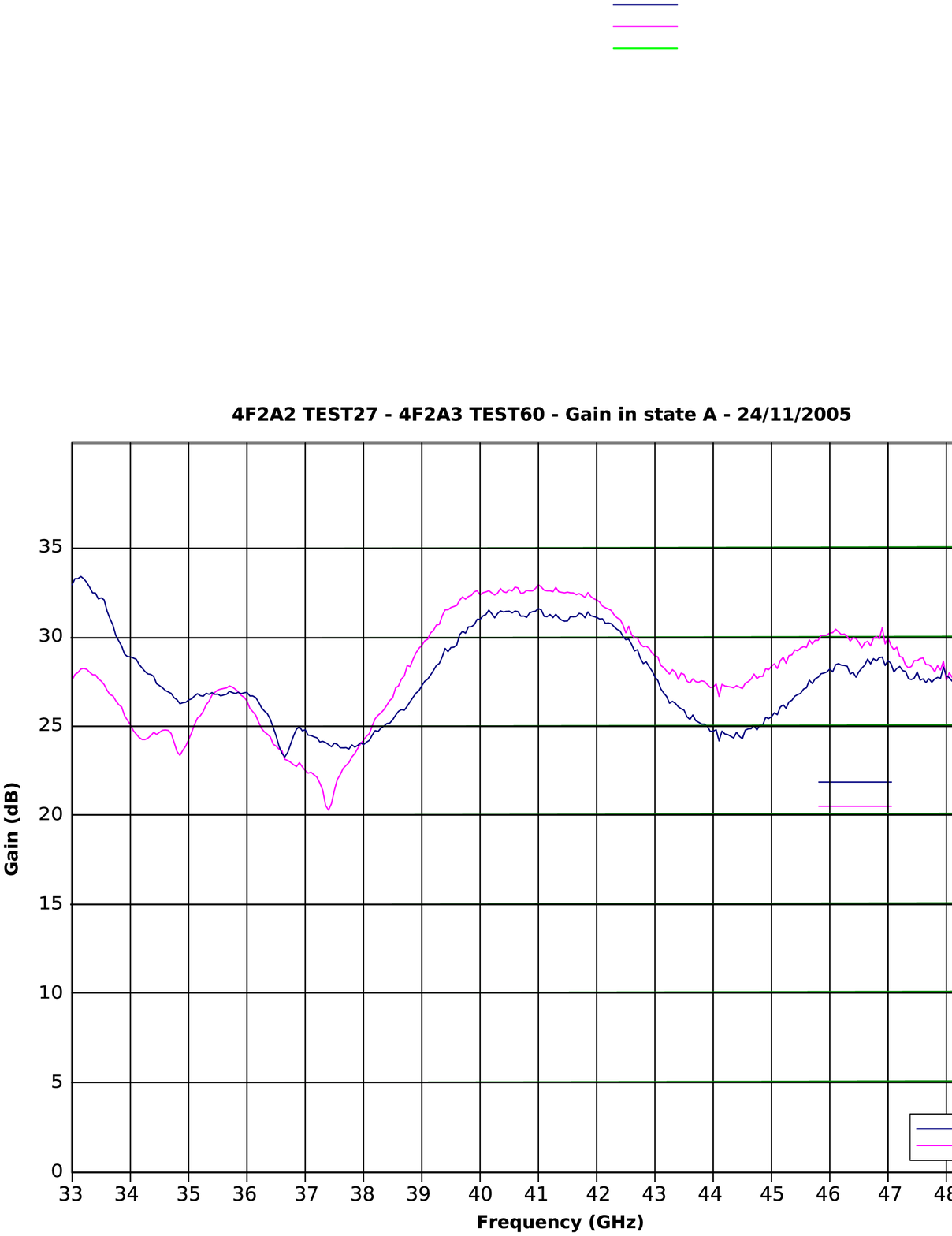}
			\caption{Gain amplitude match for two Q band LNAs}
			\label{fig:Qbandgain}
			\end{figure*}

\begin{figure*}
			\includegraphics[angle=0,width=17cm,clip]{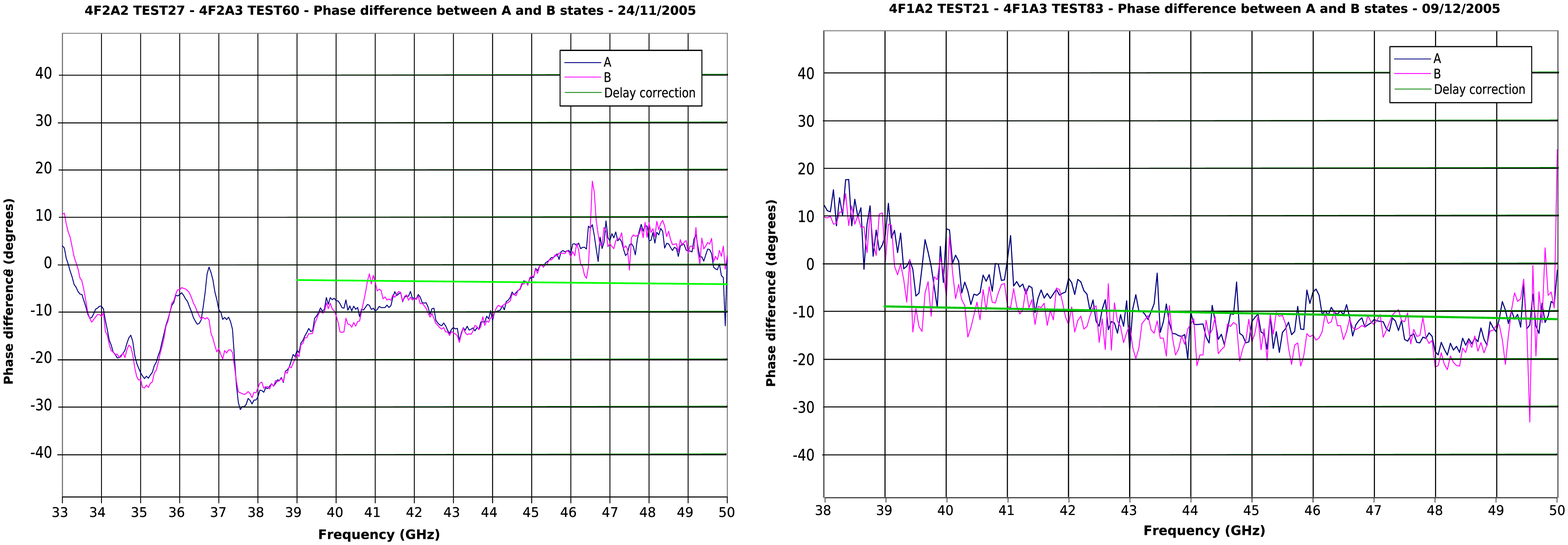}
			\caption{Phase difference for two Qband LNAs.}
			\label{fig:Qbandphase}
			\end{figure*}

			It can be seen from Figures \ref{fig:Qbandgain} and \ref{fig:Qbandphase}, that gain amplitude and phase are well matched for the two Q-band amplifiers shown.

			Finally, cold tests were repeated to ensure that the noise temperature and gain profiles were still in specification.

	\subsection{LNA performances}

		The gain and noise temperature performance across the band of each amplifier is illustrated in Figure \ref{fig:gain30} for the 30~GHz FEMs and Figure \ref{fig:gain44} for the 44~GHz FEMs. The performances of all the amplifiers in each FEM are summarised in Tables \ref{tbl:meanperformance30} and \ref{tbl:meanperformance44},  as means taken across the band.

		\begin{table}
		\caption[Summary of mean performance]{\label{tbl:meanperformance30} Summary of mean performance of 30~GHz LNAs in FEMs 3F1 and 3F2. For the definition of effective bandwidth see section \ref{sec:gainprofile}}
		\begin{center}
		\begin{tabular}{ c c c c }
		\hline \hline
		Amp \# & Mean NT & Mean Gain & Effective \\
		 & across band (K) & across band (dB) & Bandwidth (dB) \\  \hline
		3F1A1 & 6.7 & 31.8 & 4.5 \\
		3F1A2 & 7.8 & 32.1 & 4.9 \\
		3F1A3 & 7.8 & 35.6 & 5.7 \\
		3F1A4 & 8.0 & 32.5 & 5.6 \\
		3F2A1 & 9.76 & 32.13 & 5.3 \\
		3F2A4 & 11.15 & 32.60 & 5.8 \\
		3F2A2 & 6.30 & 33.51 & 5.2 \\
		3F2A3 & 7.05 & 33.76 & 5.7 \\ \hline
		\end{tabular}
		\end{center}
		\end{table}

		Figure \ref{fig:gain30} centred at 30~GHz shows noise temperatures for the amplifiers typically varying from just over 10~K at the band edges to $\sim$~5~K in the middle of the band with a gain varying from just over 35~dB to just under 30~dB.   The average values over all the 30~GHz amplifiers are 8.1~K noise temperature, 33.0~dB gain and a bandwidth of 5.3~GHz. The similar average figures for the 44~GHz amplifiers are 13.9~K noise temperature, 34.3~dB gain and 7.8~GHz bandwidth. Comparison with the values for the complete FEMs (Tables 6 and 7) show that these numbers are just a few percent better, indicating that there is very little extra loss in the FEMs.

		\begin{table}
		\caption[44~GHz LNAs, Summary of  performance]{\label{tbl:meanperformance44} Summary of mean performance of 44~GHz LNAs }
		\begin{center}
		\begin{tabular}{ c c c c } \hline \hline
		Amp \# & Mean NT & Mean Gain & Effective \\
		 & across band (K) & across band (dB) & Bandwidth (dB) \\  \hline
		4F1A1 & 14.3 & 34.5 & 7.1 \\
		4F1A4 & 14.6 & 35.8 & 7.5 \\
		4F1A2 & 13.3 & 35.1 & 8.95 \\
		4F1A3 & 15.5 & 34.5 & 7.9 \\

		4F2A1 & 14.1 & 34.0 & 7.2 \\
		4F2A4 & 14.3 & 33.3 & 8.5 \\
		4F2A2 & 11.6 & 33.0 & 8.2 \\
		4F2A3 & 13.6 & 33.9 & 8.2 \\

		4F3A1 & 13.5 & 34.9 & 7.7 \\
		4F3A4 & 13.7 & 34.4 & 7.7 \\
		4F3A2 & 13.9 & 34.0 & 6.9 \\
		4F3A3 & 14.2 & 34.2 & 7.5 \\ \hline
		\end{tabular}
		\end{center}
		\end{table}

\section{Qualification}
\label{assembly}

	Once the amplifier construction was complete, they were assembled into the FEM bodies, four per FEM (two radiometer front ends, one for each sense of polarisation).  They were then ready for qualification and performance verification.  In this section we describe the qualification of the assembled FEMs.

	\subsection{ FEM Vibration Qualification}

		A structural analysis was carried out with the assistance of the Engineering Department, University of Manchester, for both 30 and 44~GHz FEMs before commencing the vibration testing.  The critical frequencies under the Ariane V vibration spectrum ($\sim$150~Hz) were found to be very much lower frequency than would be expected for any resonances affecting the bonds. The lowest vibration natural frequencies of the FEMs were 1.8~kHz at 30~GHz and 0.92~kHz at 44~GHz. The vibration testing thus tested only the structural integrity of the FEM blocks. 

\begin{figure}
		\centering
		\includegraphics[angle=0,width=15 cm]{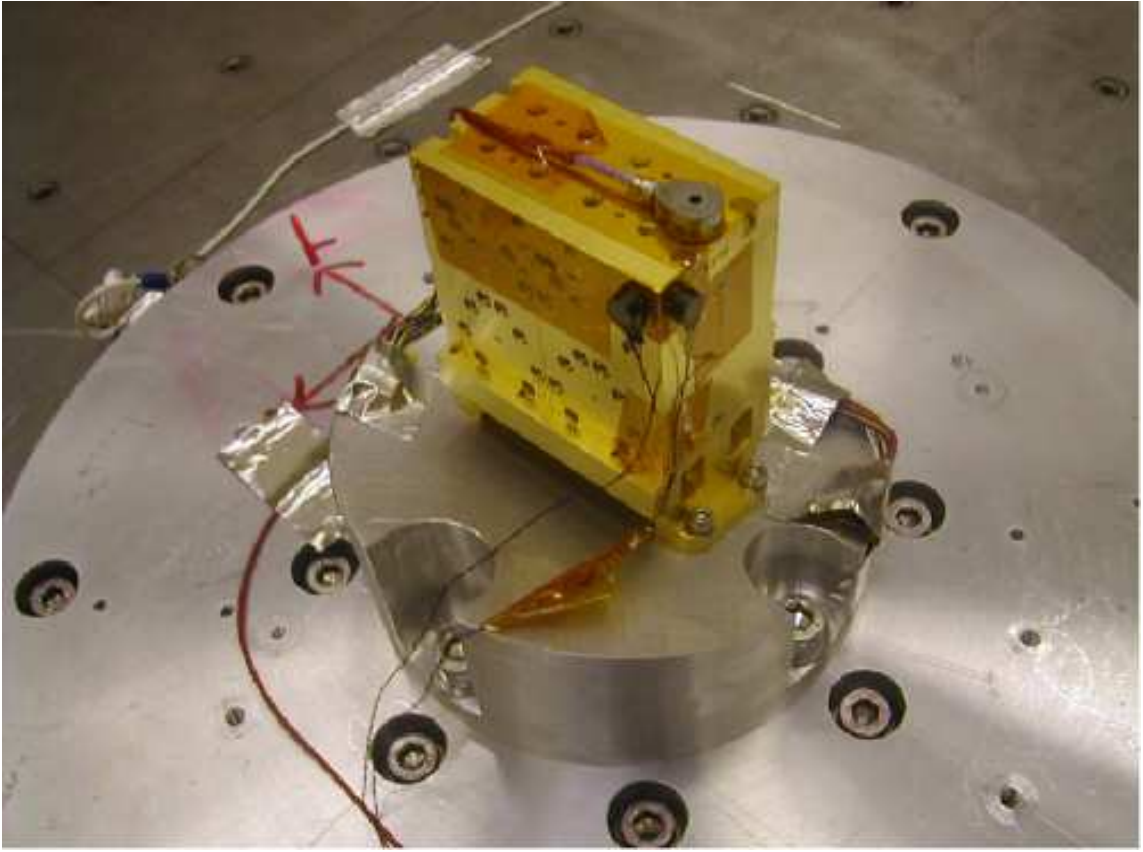}
		\caption{A Ka band FEM on the vibration table during a vibration test at Rutherford Appleton Laboratories}
		\label{fig:FEMvibration}
		\end{figure}

		The FEMs were vibrated at ESA-agreed QM and FM levels at CCLRC to qualify for flight operation. One of the Ka band FM FEMs is shown on the vibration table in Figure \ref{fig:FEMvibration}. Before departure for CCLRC, and immedately on return, room temperature S-parameter measurements made with the VNA were used to measure the RF performance.  At CCLRC, the bias settings were checked before, after and in between each vibration axis to determine that no changes were taking place during the vibration procedure.  As predicted by the structural analysis, no significant changes in either bias settings or RF performance were found for any of the FEMs.

	\subsection{Thermal Vacuum Testing}

\begin{figure}
		\centering
		\includegraphics[angle=0,width=15 cm]{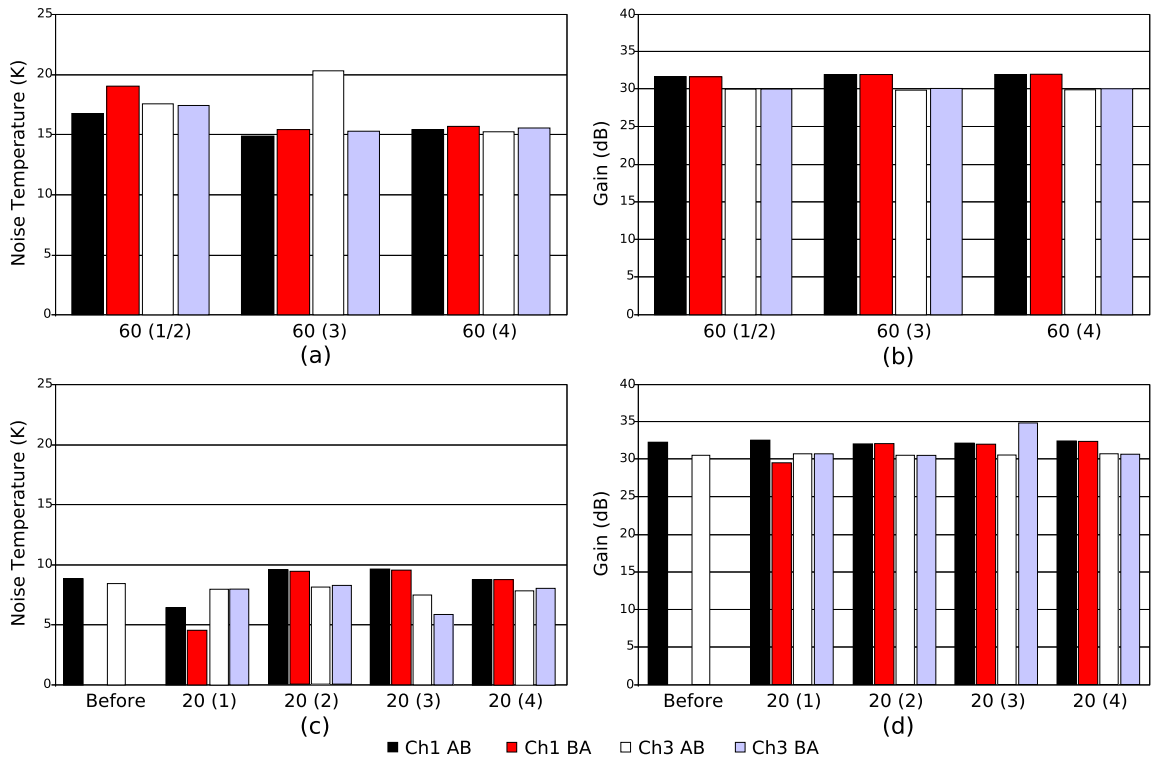}
		\caption{Thermal cycles, a) Average noise temperature over 4 cycles at 60~K, b) Average gain at 60~K, c) Average noise temperature over 4 cycles at 20~K., d) Average gain at 20~K}
		\label{fig:thermalcycles}
		\end{figure}

		 Special cryostats were built for thermal vacuum and performance testing of the 30 and 44~GHz FEMs. The sky waveguide load was controlled by a Lakeshore controller for accurate temperatures in the range 20-30~K. In this way accurate Y factors (Equation \ref{eqn:Y}) for the FEMs were measured and noise temperatures to 1~K, including systematic and random errors, were obtained. Each FEM was put through at least 4 (and in some cases, several more) cycles, over a few days, between room temperature and 20~K as prescribed by ESA, with test measurements being taken at both warm and cold stages of the cycles to confirm that the FEMs were not showing significant degradation in performance. Examples of the change in measured performance for FEM 3F1 are shown in Figure \ref{fig:thermalcycles}, the fluctuations also giving an indication of the repeatability of the results from one test to another. No significant trends were found for any FEM. 

		\subsubsection{ Long term variation}

			During the qualification and performance verification the FEMs were operated for many hours in both cryogenic and warm conditions.  All were checked for long term variation in performance. Comparison of measurements taken on 4F3 at the beginning of the thermal vacuum cycles and at the end of final performance testing (Table \ref{tbl:longterm}, see also Section \ref{performanceresults}) show that the performance was not degraded significantly over several weeks of testing and thermal cycling ($\Delta T_{n} \leq 1 K, \Delta G \leq 0.2 dB$).

			\begin{table}
			\caption[Example of long term testing at 20 K.]{\label{tbl:longterm} Example of long term testing at 20~K of the four channels. NT is the noise temperature and I is the isolation.}
			\begin{center}
			\begin{tabular}{ r c c c c c c c c } \hline \hline
			Ch & \multicolumn{2}{c}{NT pre} & \multicolumn{2}{c}{NT post} & Gain & Gain & I & I \\
			 & \multicolumn{2}{c}{(K)} & \multicolumn{2}{c}{(K)} & pre & post & pre & post \\ \hline
			 & & & (dB) & (dB) & (dB) & (dB) & (\%) & (\%) \\ \hline
			~ & AB & BA & AB & BA & ~ & ~ & ~ & ~ \\
			1 & 16.5 & 16.2 & 15.6 & 15.3 & 30.2 & 30.4 & 2.2 & 3.3 \\
			 & AA & BB & & & & & \\
			2 & 16.1 & 16.1 & 15.7 & 16.1 & 31 & 31.2 & 2.1 & 2.7 \\
			 & AB & BA & & & & & \\
			3 & 16.3 & 16.5 & 15.4 & 15.5 & 31.5 & 31.7 & 1.5 & 1.9 \\
			 & AA & BB & & & & & \\
			4 & 15.5 & 15.4 & 15.5 & 15.4 & 32.2 & 32.1 & 3.3 & 3.7 \\ \hline
			\end{tabular}
			\end{center}
			\end{table}

	\subsection{Bias Supply}

		The bias supply used to test the FEMs had to be extremely stable and connected in such a way that no earth loops were formed.  This was achieved  by running the FEMs off trickle-charged batteries.  The space qualified bias supply in the LFI is housed in the DAE (Data Acquisition Electronics) box, and can be controlled and monitored when in space.  The following sections describe the response of the FEM to various changes in the bias supply.

	\subsection{Sensitivity to drain voltage -- operating point}

\begin{figure}
		\centering
		\includegraphics[angle=0,width=15 cm]{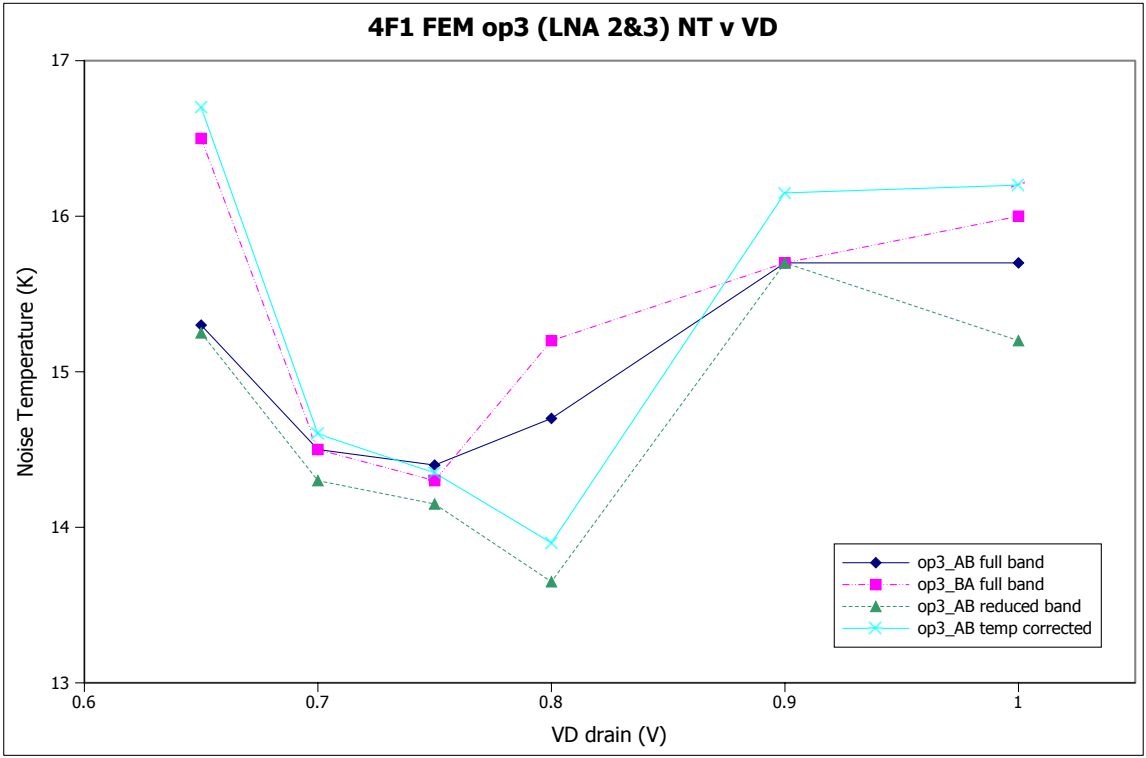}
		\caption{Variation of noise temperature with $Vd$, FEM 4F1, output 3}
		\label{fig:variationNT}
		\end{figure}

\begin{figure}
		\centering
		\includegraphics[angle=0,width=15 cm]{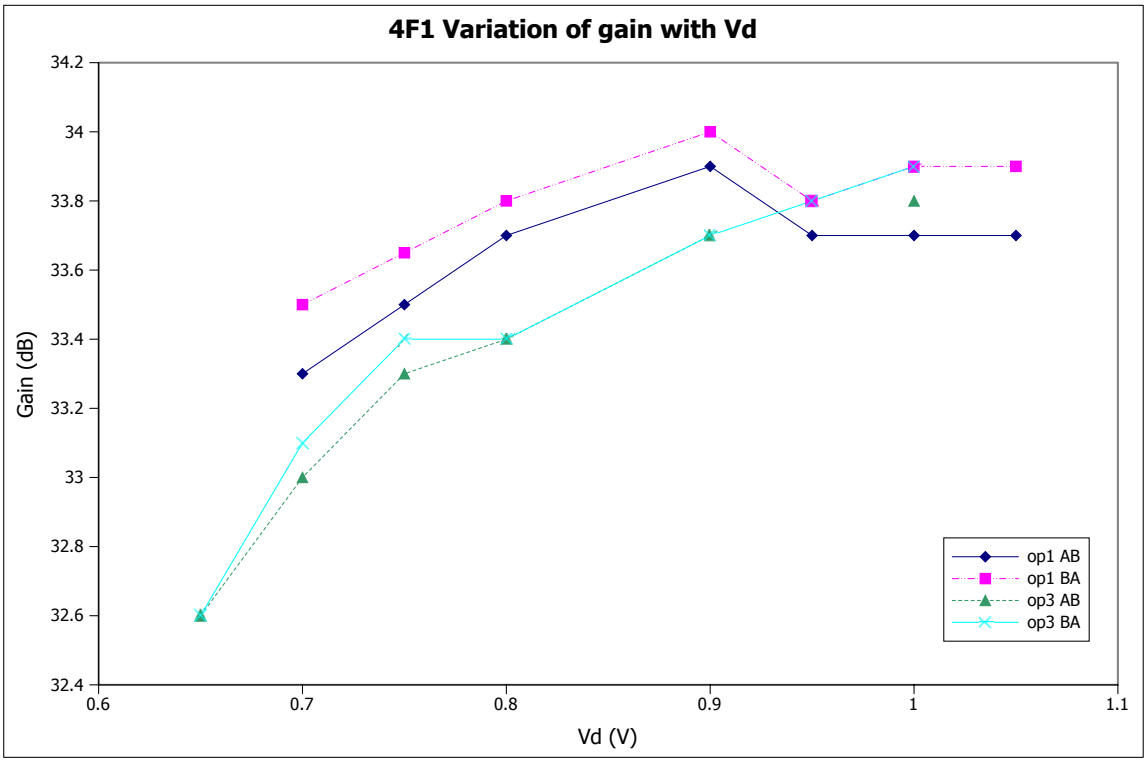}
		\caption{Variation of gain with $Vd$, FEM 4F1, output 3}
		\label{fig:variationGain}
		\end{figure}

		The dependence of noise temperature and gain on the LNA drain voltage, $Vd$, were investigated as part of the process of tuning the FEM (results for FEM 4F1 shown in Figures \ref{fig:variationNT} and \ref{fig:variationGain}). The results for 4F1 were typical of the sort of behaviour found. $Vd$ was the only parameter changed; the gate voltage, $Vg$, values being kept constant at 1.2~V and not modified to keep the drain current, $Id$, constant.  The gain increased as $Vd$ is increased up to about 1~V, then flattened off.  In this case the noise temperature measurement  showed a broad minimum between 0.7 and  0.85~V.  0.85~V  was adopted as the best compromise operating point.  The effective bandwidth was also close to maximum at this $Vd$ value.

	\subsection{ FEM sensitivity to voltage supply -- EMC testing}

		The purpose of this test is to check the sensitivity to ripple on the supply lines.The sensitivity of the FEM to drain and gate voltage was checked. Test bias break-out connections were provided for access to the raw drain voltage supplies to two LNAs. These allowed monitoring of the applied bias voltage and/or injection of modulating signals to simulate conducted RFI.

		For AC test measurements the waveform generator was set to give a sine wave modulation on the LNA drain voltage such that the modulation on the BEM output voltage could be clearly seen and measured using the spectrum analyser. For each test frequency the input modulation voltage $V_\mathrm{in}(rms)$, the BEM DC output voltage, and BEM output frequency component, $V_\mathrm{BEM} (rms)$ were recorded. The transfer function was then computed as follows:

		\begin{equation}
		F_{transfer} = \frac{V_\mathrm{BEM}(rms)}{V_\mathrm{in} (rms)\times V_\mathrm{BEM}(DC)},
		\label{eqn:TVBEM}
		\end{equation}
		
		i.e. the fractional BEM voltage variation per unit input voltage variation.
		
		The data show a decreasing response with frequency indicating that at frequencies above 100~kHz the supply lines will not affect the FEM.

\section{FEM and Radiometer Performance}
\label{performanceresults}

	The performance of each FEM, comprising two front-end radiometer sections, is characterised in terms of the following parameters: noise temperature, gain versus frequency, effective bandwidth, isolation, and  1/f knee frequency.

	FEM system tests were performed alone, and in conjunction with representative waveguides and a representative BEM. The BEMs used were models supplied by the manufacturers of the flight BEMs, and further details can be found in \citet{2009_LFI_cal_R9}.

	\subsection{ S-parameter measurements, FEMs only}

\begin{figure}
		\centering
		\includegraphics[angle=0,width=15 cm]{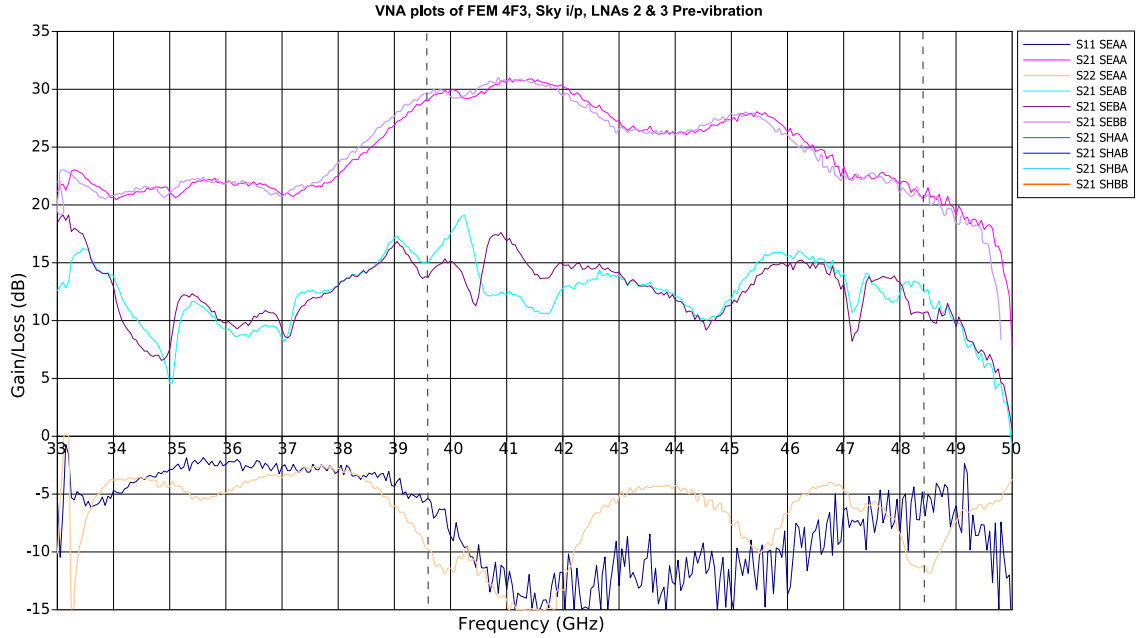}
		\caption{S-parameters as a function of frequency for 44~GHz FEM 4F3, with the input signal on the sky port, for the radiometer containing LNAs 4F3A2 and 4F3A3.  The VNA frequency range is much wider than the Planck bandwidth, 39.6 to $48.4$~GHz, indicated by dashed vertical lines.}
		\label{fig:sparametersfreq}
		\end{figure}

		Several other tests were necessary to establish FEM functionality.  The first test performed, for room temperature characterisation of the FEM, and to provide a baseline against which the FEM could be compared throughout the test sequence, was the measurement of the room temperature S-parameters. All combinations of sky and reference port input, E- and H-plane output and phase switch state for each FEM were characterised using the vector network analyser.  A restricted set of combinations were repeated whenever confirmation was required that no changes had taken place after events such as the vibration tests.  An example of these results, for FEM 4F3, is shown in Figure \ref{fig:sparametersfreq}.

		When the two phase switches, in a given radiometer, are in the same state (AA or BB), S21 indicates the forward gain state. When the phase switches are in antiphase (AB or BA), the S21 parameter gives the ``isolation state'', and the difference between the two states gives a measurement of the isolation as a function of frequency. Over the Planck band this is from 10 to 15~dB.  S11, the input return loss, is between -5 and -15~dB across the Planck band, and S22, the output return loss, is between -4 and -15~dB.

	\subsection{FEM Noise temperature and gain}

		The noise temperature of the FEM is a critical parameter, as it must be low enough to allow the instrument to detect fluctuations of the order of $10^{-5}$ to $10^{-6}$~K in the observed temperature of the cosmic microwave background as the spacecraft and its antenna beams rotate.  The noise temperature of the LNAs is dependent on the physical temperature of the devices, hence all measurements of FEM performance are made with the FEM maintained close to the Planck operating temperature of 20~K

		The receiver noise temperature is calculated according to the classical Y-factor method, whereby the FEM output is measured at two input temperatures.  The reference load is kept at a stable low temperature ($\sim$~17~K).  Within the cryostat, the sky horn is replaced by a temperature controlled waveguide load, which is varied between $T_{low} $ ($\sim$26~K) and $T_{high} $ ($\sim$50~K).  The measured voltage is proportional to $(T_{n} +T_{sky} )$, hence the ratio between the output voltages viewing hot and cold loads, or Y factor, is 

		\begin{equation}
		Y=\frac{V_\mathrm{high}}{V_\mathrm{low}} =\frac{(T_\mathrm{n} +T_\mathrm{high})}{(T_\mathrm{n}+T_\mathrm{low})}
		\label{eqn:Y}
		\end{equation}

		The gain must be such as to maximise the amplification of the signal, but must not be such that  the signal at the FEM output is enough to cause operation in the non-linear response region in the BEM. The gain versus frequency response of the FEM is also an important element in determining the effective bandwidth, although the final bandwidth is set by a filter in the BEM (see Section \ref{sec:effBW}).

		\subsubsection{Gain and noise temperature as a function of frequency}
		\label{sec:gainNTisolation}

		\begin{figure}
		\centering
		\includegraphics[angle=0,width=15 cm]{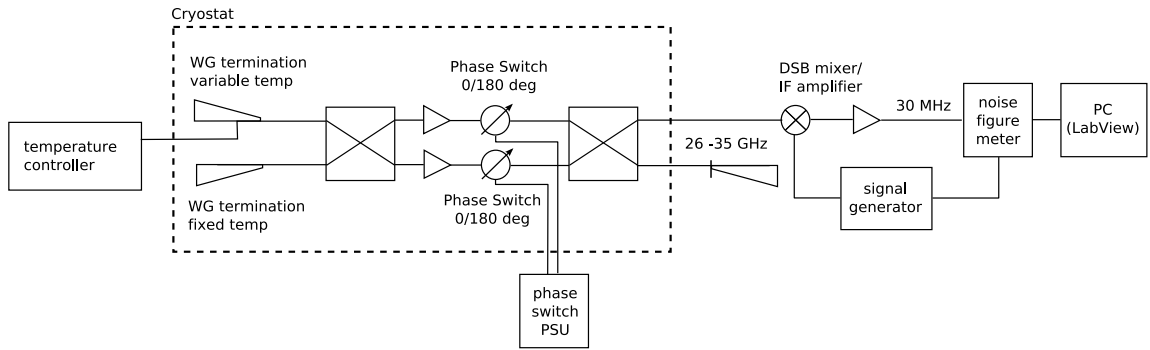}
		\caption{Measurement set up for noise temperature and gain determination as a function of frequency.  The dashed section indicates components within the cryostat, which are also the components encapsulated within the FEMs.}
		\label{fig:NTfreq}
		\end{figure}

		The gain and  noise temperature were measured at intervals across the Planck band, to define the frequency response and allow calculation of the effective bandwidth (see Section \ref{sec:effBW}). Figure \ref{fig:NTfreq} shows the set up used for each half-FEM noise temperature test. Only one FEM output at a time could be measured. The biases used were those determined as optimal in the individual LNA tests and fixed during the commoning of the gate and drain biases, and the mixer system was calibrated using an Agilent noise diode type R347B.  The reference load temperature was controlled to a fixed temperature of about 17~K. The sky input was set to a controlled temperature of about 26~K (see the description of the determination of the appropriate temperature in Section \ref{sec:linearity}). The measurement was made for each phase switch state, and the total power output from the FEM was measured across the band in steps of 100~MHz.  The sky input was then set to 50~K, the system allowed to equilibrate, and the sequence of measurements repeated.

\begin{figure}
		\centering
		\includegraphics[angle=0,width=15 cm]{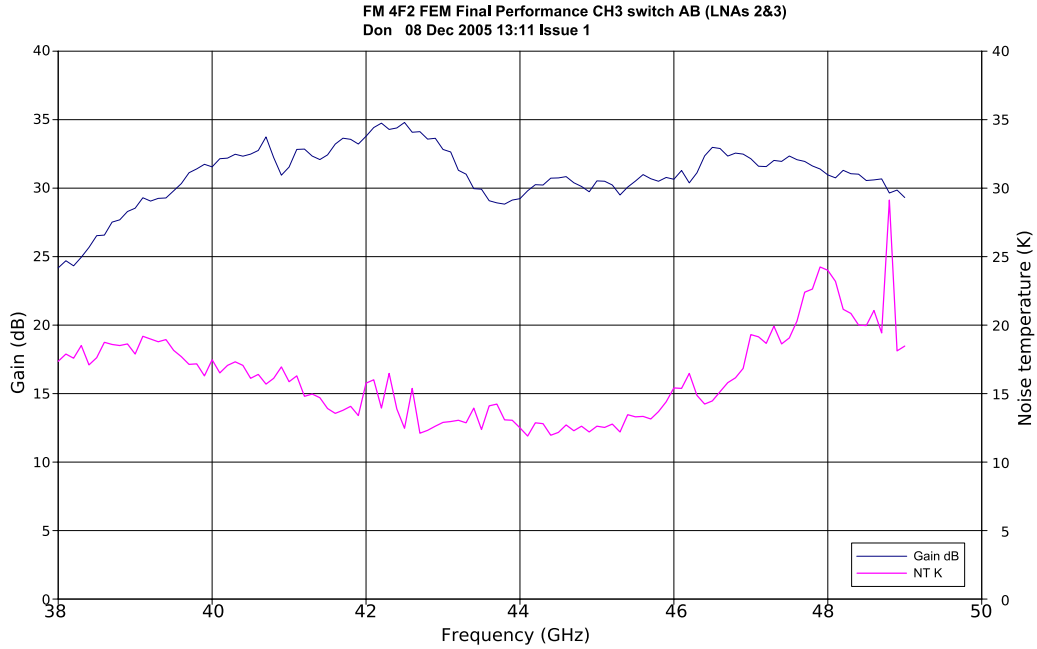}
		\caption{Gain and noise temperature as a function of frequency across the Planck band for 44~GHz FEM 4F2, Channel 3.}
		\label{fig:NTfreq44}
		\end{figure}

		An example of the variation of gain and noise temperature as a function of frequency is shown for FEM 4F2 in Figure \ref{fig:NTfreq44}. Tables \ref{tbl:30flightperformance} and \ref{tbl:44flightperformance} summarise the results for all the FEMs at both frequencies, which are given as band averages over the nominal Planck band. The results of all the high gain states are presented, for each radiometer, together with the mean for each radiometer. Both internal radiometer names (e.g. 3F1) and the LFI numbers used in radiometer assembly and testing are given, so cross comparison can be made with the results obtained from the measurements of the noise properties of the flight receivers \citep{2009_LFI_cal_R2,2009_LFI_cal_M3}.

		For all these noise temperature measurements, the measurement uncertainty is estimated at  1~K.  At 30~GHz, for RCA27 (3F2), both channels are lower than the noise temperature requirement of 8.6~K (cf. Table \ref{tbl:specifications}).  For RCA28 the noise temperatures are 1-2~K higher. At 44~GHz, the noise temperatures are fairly consistent for all the FEMs. The performances achieved are highly satisfactory, and represented the state of the art at the time when the amplifiers were built.

		The mean gains at 30~GHz, $\sim$~31~dB, do not formally meet the requirements, but in practice, given the actual realisation of the BEMs, higher gain was not wanted since the signal level at the BEM would have caused saturation had the gain approached the formal requirement.

		A similar comment applies to the gains at 44~GHz, though the gains were intrinsically capable of being higher here because the amplifiers had five stages rather than four. Low noise temperature was regarded as the primary target in the LNA tuning procedure. In the case of the 44~GHz, the complex iteration process to balance the FEM and BEM gain lead to an overall radiometer gain higher than optimal resulting in a slight residual non-linearity in the radiometer response \citep{2009_LFI_cal_M4,2009_LFI_cal_R4}.

		Tables \ref{tbl:30flightperformance} and \ref{tbl:44flightperformance} also contain details of the effective bandwidth and isolation which were derived from the present measurements, and will be described in more detail in Section \ref{sec:effBW}.

		\begin{table*}
		\caption[]{\label{tbl:30flightperformance} 30~GHz flight models performances, measured with the FEMs at 20~K physical temperature.}
		\begin{center}
		\begin{tabular}{ c c c c c c c c } \hline \hline
		FEM & Ch & Phase switch & Noise & Gain & Effective & Isolation & NT (K) Isolation \\
		 &  & state & temp (K) & (dB) & bandwidth (GHz) & (\%) & Corrected \\ \hline 
		RCA28 & 1 & AB & 10.6 & 32 & 4.9 & 6.1 & 8.9 \\
		(3F1A2, 3F1A3) & & BA & 10.6 & 32 & 4.9 & ~ & ~ \\ 
		(M2, M1) & 2 & AA & 10.7 & 31.9 & 5.1 & 4.2 & 9.6 \\
		~ & ~ & BB & 10.7 & 31.8 & 5.1 & ~ & ~ \\ \hline
		\multicolumn{3}{r}{Mean, radiometer 1} & 10.7 & 31.9 & 5.0 & 5.2 & 9.3 \\ \hline
		RCA28 & 3 & AB & 9.8 & 30.2 & 5.1 & 4.4 & 8.5 \\
		(3F1A1, 3F1A4) &  & BA & 9.8 & 30.2 & 5.1 & ~ & ~ \\
		(S2, S1) & 4 & AA & ~ & ~ & ~ & 3.9 & 8.9 \\
		~ & ~ & BB & 10 & 30.3 & 5.2 & ~ & ~ \\ \hline
		\multicolumn{3}{r}{Mean, radiometer 2} & 9.9 & 30.2 & 5.1 & 4.2 & 8.7 \\ \hline
		RCA27 & 1  & AB & 7.2 & 31.9 & 6.1 & 2.9 & ~ \\
		(3F2A1,3F2A4) &  & BA & 7.2 & 31.9 & ~ & ~ & ~ \\
		(S2, S1) & 2 & AA & 6.9 & 32.2 & 6.3 & 3.6 & ~ \\
		~ & ~ & BB & 7.3 & 32.2 & ~ & ~ & ~ \\ \hline
		\multicolumn{3}{r}{Mean, radiometer 1} & 7.2 & 32.1 & 6.2 & 3.3 & ~ \\ \hline
		RCA27 & 3 & AB & 8.1 & 31 & 6.5 & 3.3 & ~ \\
		(3F2A2,3F2A3) &  & BA & 8 & 31 & ~ & ~ & ~ \\
		(M2, M1) & 4 & AA & 7.7 & 31.2 & 6.4 & 3 & ~ \\
		~ & ~ & BB & 7.9 & 31.2 & ~ & ~ & ~ \\ \hline
		\multicolumn{3}{r}{Mean, radiometer 2} & 7.9 & 31.1 & 6.5 & 3.2 & ~ \\ \hline
		\end{tabular}
		\end{center}
		\end{table*}

		\begin{table*}
		\caption[]{\label{tbl:44flightperformance} 44~GHz Flight Model performances, measured at 20~K physical temperature}
		\begin{center}
		\begin{tabular}{ c c c c c c c c } \hline \hline
		FEM & Ch& Phase switch & Noise & Gain & Effective & Isolation & NT (K) Isolation \\
		 & & state & temp (K) & (dB) & Bandwidth (GHz) & (\%) & Corrected \\ \hline 
		RCA25 & 1 & AB & 15.4 & 33.7 & 6.9 & 3 & 14.8 \\ 
		(4F1A1, 4F1A4) &  & BA & 15.3 & 33.9 & 6.7 & ~ & ~ \\ 
		(M1, M2) & 2 & AA & 15.6 & 34.6 & 6.9 & 3.6 & 14.7 \\ 
		~ & ~ & BB & 15.4 & 34.3 & 7 & ~ & ~ \\ \hline
		\multicolumn{3}{r}{Mean, radiometer 1} & 15.4 & 34.1 & 6.9 & 3.3 & 14.8 \\ \hline
		RCA25 & 3 & AB & 15.4 & 33.4 & 7.9 & 1.8 & 15.1 \\ 
		(4F1A2, 4F1A3) &  & BA & 15 & 33.5 & 7.9 & ~ & ~ \\ 
		(S2, S1) & 4 & AA & 14.4 & 33.9 & 7.7 & 0.7 & 14.3 \\ 
		~ & ~ & BB & 14.5 & 33.7 & 7.7 & ~ & ~ \\ \hline
		\multicolumn{3}{r}{Mean, radiometer 2} & 14.8 & 33.6 & 7.8 & 1.3 & 14.7 \\ \hline
		RCA24 & 1 & AB & 15.5 & 30.6 & 7.2 & 4.9 & 13.9 \\ 
		(4F2A1, 4F2A4) &  & BA & 15.4 & 30.5 & 7.3 & ~ & ~ \\ 
		(M2, M1) & 2 & AA & 15.6 & 31.3 & 7.1 & 3.8 & 14.5 \\ 
		~ & ~ & BB & 15.4 & 31.1 & 7.1 & ~ & ~ \\ \hline
		\multicolumn{3}{r}{Mean, radiometer 1} & 15.5 & 30.9 & 7.2 & 4.4 & 14.2 \\ \hline
		RCA24 & 3 & AB & 15.2 & 31.8 & 7.9 & 3.2 & 14.3 \\ 
		(4F2A2, 4F2A3) & & BA & 15.3 & 31.8 & 7.8 & ~ & ~ \\ 
		(S1, S2) & 4 & AA & 15.6 & 32.4 & 7.8 & 3.8 & 14.3 \\ 
		~ & ~ & BB & 15.5 & 32.1 & 7.8 & ~ & ~ \\ \hline
		\multicolumn{3}{r}{Mean, radiometer 2} & 15.4 & 32.0 & 7.8 & 3.5 & 14.3 \\ \hline
		RCA26 & 1 & AB & 15.4 & 30.7 & 7.07 & 5.1 & 13.5 \\ 
		(4F3A2, 4F3A3) &  & BA & 15.2 & 30.8 & 7.04 & ~ & ~ \\ 
		(S2, S1) & 2 & AA & 14.7 & 31.5 & 7.17 & 5.9 & 13 \\ 
		~ & ~ & BB & 15.5 & 31.1 & 7.23 & ~ & ~ \\ \hline
		\multicolumn{3}{r}{Mean, radiometer 1} & 15.2 & 31.0 & 7.1 & 5.5 & 13.3 \\ \hline
		RCA26 & 3 & AB & 17.9 & 32.0 & 6.77 & 6.4 & 15.6 \\ 
		(4F3A1, 4F3A4) &  & BA & 18.1 & 32.0 & 6.84 & ~ & ~ \\ 
		(M1, M2) & 4 & AA & 16.7 & 32.6 & 6.65 & 7.1 & 13.9 \\ 
		~ & ~ & BB & 16.1 & 32.5 & 6.63 & ~ & ~ \\ \hline
		\multicolumn{3}{r}{Mean, radiometer 2} & 17.2 & 32.3 & 6.7 & 6.8 & 14.8 \\ \hline 
		\end{tabular}
		\end{center}
		\end{table*}

		\subsubsection{ Temperature Susceptibility}

\begin{figure}
			\centering
			\includegraphics[angle=0,width=15 cm]{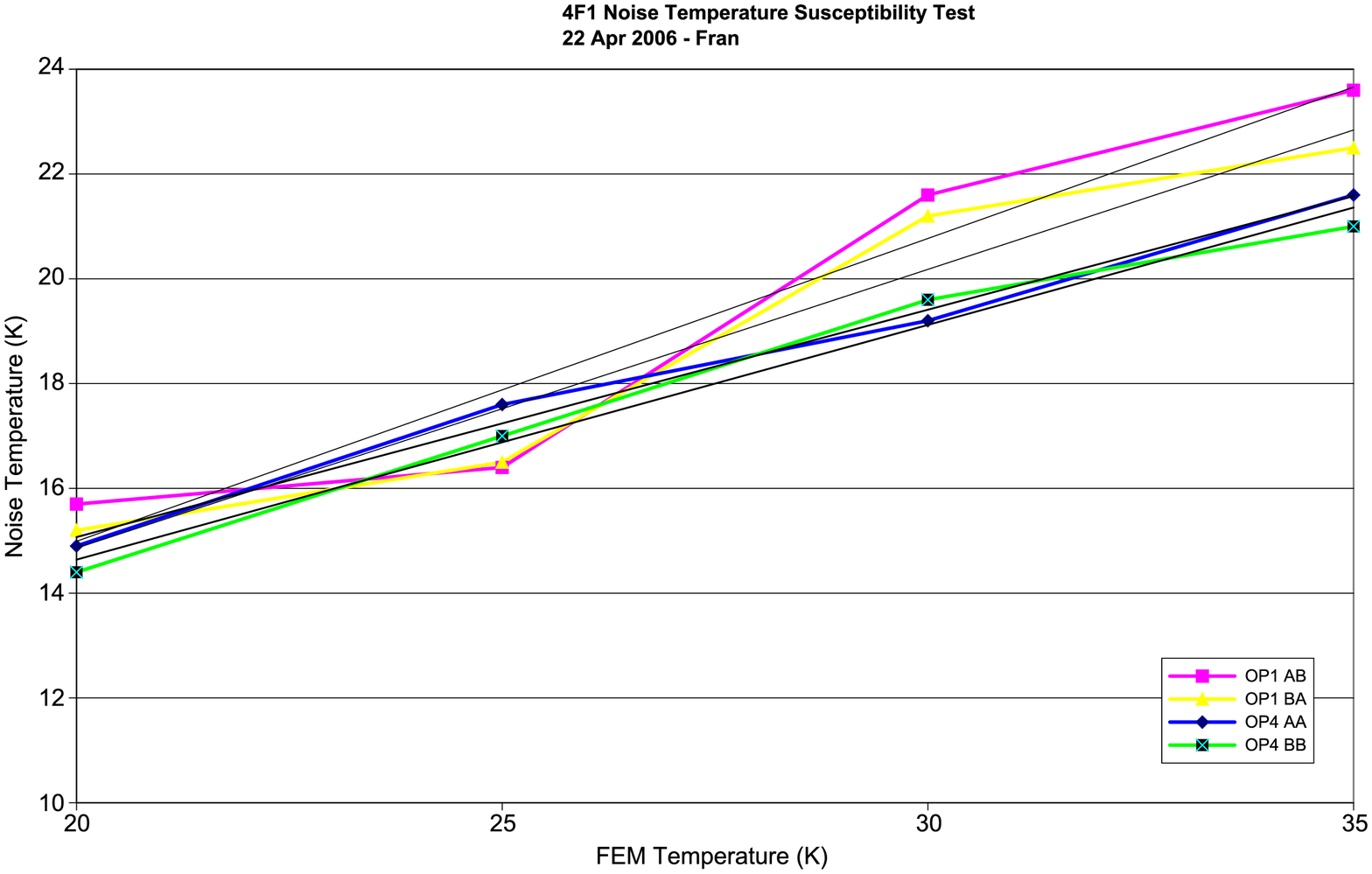}
			\caption{Noise temperature susceptibility to FEM physical temperature, 4F1}
			\label{fig:NT4F1}
			\end{figure}

\begin{figure}
			\centering
			\includegraphics[angle=0,width=15 cm]{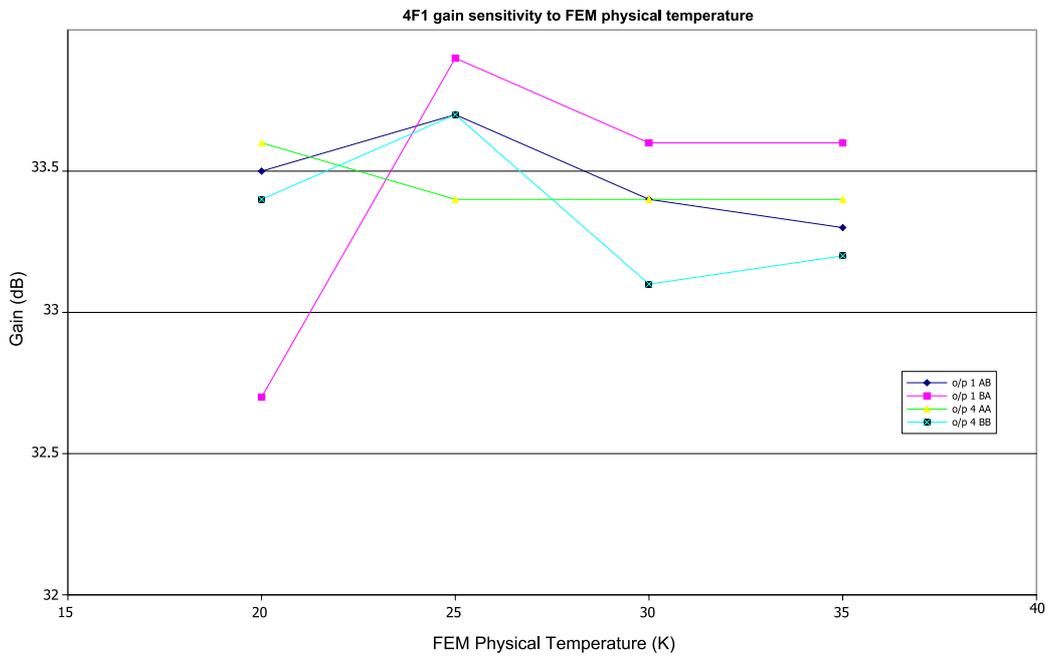}
			\caption{Gain susceptibility to FEM physical temperature, 4F1}
			\label{fig:Gain4F1}
			\end{figure}

			Both noise temperature and gain vary with the physical temperature of the FEM, because the properties of the HEMT devices are still dependent on physical temperature at 20~K.  The experimentally observed dependence of noise temperature $T_\mathrm{n}$ of InP HEMTS vs ambient temperature $T_\mathrm{a} $ approximately decreases linearly with $T_\mathrm{a}$  between 300 and 80~K while  below 80~K  approximately decreases with $\sqrt{T_\mathrm{a} }$ \citep{1994Pospieszalski,2005Pospieszalski}. In fact the data in Figure \ref{fig:NT4F1} show the dependence of noise temperature approximately proportional to $\sqrt{T_\mathrm{a} }$. The variation of gain and noise temperature with FEM(4F1) physical temperature found for the CRYO3 2060 devices are shown in Figures \ref{fig:NT4F1} and \ref{fig:Gain4F1}. The FEM cryostat could not be taken much below 20~K, but for this FEM if the behaviour for 4F1 continues to 4~K physical temperature, extrapolation would have taken the noise temperature from $\sim$~15~K to $\sim$~6~K.

			The gain susceptibility is less clear, but the suggestion is that the peak gain is close to being achieved at around 20-25~K, as shown in Figure \ref{fig:Gain4F1}.

	\subsection{Effective bandwidth}
	\label{sec:effBW}

		\subsubsection{Integration over the gain profile}
		\label{sec:gainprofile}

			The effective bandwidth, which is required to be 20~\% of the central observing frequency at each radiometer frequency (see Table \ref{tbl:specifications}), is defined as:

			\begin{equation}
			\beta =\frac{\left[\int G(\nu )d\nu  \right]}{\int \left[G(\nu )^{2} \right]d\nu  } ^{2},
			\label{eqn:beta}
			\end{equation}

			or practically for discrete measurements:

			\begin{equation}
			\beta =\Delta f.\frac{N}{N+1} .\frac{\left(\sum _{i=1}^{N}V_\mathrm{out} (i) \right)^{2} }{\sum _{i=1}^{N}\left(V_\mathrm{out} (i)\right)^{2}},
			\label{eqn:beta2}
			\end{equation}

			where $N$ is the number of frequency points, $\Delta $f is the frequency step, and $V_{out}(i)$ is the output voltage at each frequency. 

			This bandwidth can be determined directly by calculation using the data obtained during the gain measurements for the FEMs described in Section \ref{sec:gainNTisolation} above. The calculations for the FEM alone quoted in Table \ref{tbl:30flightperformance} and Table \ref{tbl:44flightperformance} have been carried out over the appropriate nominal Planck bandwidth, 27-33~GHz for central frequency 30~GHz, and 39.6-48.4~GHz for central frequency 44~GHz.  Such calculations would be expected to yield lower values than would be measured for a complete radiometer chain, because the system as realized will not cut off sharply at the nominal frequencies. At 30~GHz, FEM 3F2 meets the requirement (6~GHz), and 3F1 is slightly below, while at 44~GHz the bandwidths are all slightly low.  These values are indicative but are not characteristic of the flight radiometer bandwidths.  Comparison should be made with the effective bandwidth measurements made for the flight radiometer chain assemblies (RCAs) and during the RAA tests \citep{2009_LFI_cal_M3,2009_LFI_cal_M4}.

		\subsubsection{Noise equivalent bandwidth from the radiometer data}

			The FEM by itself does not define the radiometer bandwidth, since there is no filter in the FEM, and the intrinsic FEM band is therefore much wider band than the specification.  Assuming the feed horn, OMT and waveguide bandpasses are not restrictive, the radiometer bandwidth depends mainly on the combination of the FEM amplifier band and the MMIC amplifier band, and the filter and detector in the BEM.  Hence, it is strictly only meaningful to measure the combination of FEM and BEM, together with the interconnecting waveguides.  

			The white noise level $\Delta V$($\propto \Delta T$) sets the fundamental limit on the signal level which can be measured, and the ratio $\Delta V/V$ gives the primary measurement of the sensitivity of the radiometer, independent of any detector offset.  Both these quantities can be determined from the noise characteristics of the radiometer time-ordered data streams.  The radiometer bandwidth is given by the well-known radiometer relationship:

			\begin{equation}
			\frac{\Delta T}{T} =\frac{1}{\sqrt{\beta \tau } },
			\label{eqn:dToverT}
			\end{equation}

			where  $\Delta T$ is the minimum detectable temperature change, $T$ is the overall system noise temperature, $\beta$ is the pre-detection bandwidth in Hz (assumed to be a tophat), and $\tau$ is the integration time in seconds. The rms white noise and mean noise temperature were determined from the time-ordered data streams (see Section \ref{sec:operationmode}) and the effective bandwidth derived using:

			\begin{equation}
			\beta =\frac{1}{(\Delta V/V)^{2} } .\frac{400}{77},
			\label{eqn:beta3}
			\end{equation}

			where $\Delta V$ is the rms white noise level in $V/\sqrt{Hz}$, $V$ is the average voltage level of the time series and is a measure of the signal power, and the integration time is effectively 1 second since the oversampled data are averaged to give an effective 100\% duty cycle.

			The results from two switched data streams taken with FEM 3F2 are shown in Table \ref{tbl:noiseequivalentBW}. The factor 400/77 applied in the final calculation of the bandwidth is to correct for the fact that 23\% of the data at the edge of each square wave was blanked to remove switching transients (100/77), and to account for the data averaging mentioned above (one factor of $\sqrt{2}$) and the differencing of the two channels (other factor of $\sqrt{2}$).

			\begin{table}
			\caption[]{\label{tbl:noiseequivalentBW} Noise equivalent bandwidth determined from time-ordered data stream obtained with radiometer 3F2  using first the channel containing amplifiers 3F2A1 AND 3F2A4, then the channel containing amplifiers 3F2A2 and 3F2A3.}
			\begin{center}
			\begin{tabular}{ c c c } \hline \hline
			Channel & 3F2A1 and 3F2A4 & 3F2A2 and 3F2A3 \\
			 & $1/(\Delta V/V)^{2}\frac{400}{77}$ GHz & $1/(\Delta V/V)^{2}.\frac{400}{77}$ GHz \\ \hline 
			0 & 4.30 & 4.35 \\
			1 & 5.20 & 5.16 \\
			2 & 5.40 & 6.60 \\
			3 & 5.62 & 5.67 \\ \hline 
			\end{tabular}
			\end{center}
			\end{table}

			Inevitably, the result will vary from BEM to BEM, so the results quoted in Table \ref{tbl:noiseequivalentBW}, which are lower than the bandwidths measured by band integration for the FEM alone, are only relevant to the specific QM representative BEMs used. The flight performance is determined by the FEMs in combination with the flight BEMs \citep[][]{2009_LFI_cal_R9} and other RCA elements.

	\subsection{Isolation}

		If a voltage change is introduced in one channel of the radiometer, the isolation is the ratio of the resulting change in the voltages of the other channel to the known introduced change.

		The basic definition of isolation, L, is given by:

		\begin{equation}
		L=\frac{V_\mathrm{fixed2} -V_\mathrm{fixed1} }{V_\mathrm{change2} -V_\mathrm{change1} },
		\label{eqn:L}
		\end{equation}

		where $V_{fixed}$ refers to the voltages in the nominally constant channel before and after the change, and $V_{change}$ refers to the voltages in the channel where the change was introduced. These are measures of the signal power. $L=0$ gives full isolation and $L=1$ gives full coupling.

		This definition does not take into consideration the fact that in the JBO test cryostat, part of the rise seen in the nominally ``fixed'' load is due to parasitic heating of the load, as the fixed and variable loads can not be sufficiently thermally separated to eliminate any fixed load heating.  The isolation determined in this way will overestimate the effect on the ``fixed'' load.  

		Including the parasitic heating effect, the isolation can be expressed as:

		\begin{equation}
		L=\frac{(\Delta V_\mathrm{fixed} -C\Delta T_\mathrm{fixed} )}{(\Delta V_\mathrm{fixed} -C\Delta T_\mathrm{fixed} +\Delta V_\mathrm{change} )},
		\label{eqn:L2}
		\end{equation}

		where:

		$\Delta V_{fixed}=V_{fixed2} -V_{fixed1}$,
		
		$\Delta V_{change}=V_{change2} -V_{change1}$,
		
		$\Delta T_{fixed}=T_{fixed2} -T_{fixed1}$,
		
		$\Delta T_{change}=T_{change2} -T_{change1}$,
		
		and $C$ is the temperature to voltage conversion factor (see Section \ref{sec:linearity}). This definition still does not include the effect of the slight change in FEM temperature induced by raising the load temperature, but these are negligible compared to the parasitic load heating effect.

		\subsubsection{Isolation as a function of frequency}
		\label{sec:isolationfreq}
		
			In order to derive the isolation as a function of frequency across the LFI band, a modification is required.  For a given phase switch state and a given FEM output the Y factor for one channel can be calculated as a function of frequency from:

			\begin{equation}
			Y=\frac{V_\mathrm{Hot} }{V_\mathrm{Cold} },
			\label{eqn:Y2}
			\end{equation}

			where the voltages are measured with the variable load hot and cold respectively.

			The isolation as a function of frequency is then given by:

			\begin{equation}
			L =\frac{(T_\mathrm{C1} (Y1-1)+T_\mathrm{C2} Y2(1-Y1)+(1-Y2)(T_\mathrm{H1} -T_\mathrm{H2} Y1))}{(T_\mathrm{C1} (Y1+Y2-2)+T_\mathrm{C2} (Y2-Y1(2Y2-1))-T_\mathrm{H1} (Y1+Y2-2)+T_\mathrm{H2} (Y1(2Y2-1)-Y2))}
			\label{eqn:isolation}
			\end{equation}
			
			where:

			$T_\mathrm{H1}=$ higher variable load temperature ($\sim$~50 K),

			$T_\mathrm{H2}=$ lower variable load temperature ($\sim$~20 K),

			$T_\mathrm{C1}=$ higher fixed load temperature ($\sim$~17 K),

			$T_\mathrm{C2}=$ lower fixed load temperature ($\sim$~17 K).

\begin{figure}
			\centering
			\includegraphics[angle=0,width=15 cm]{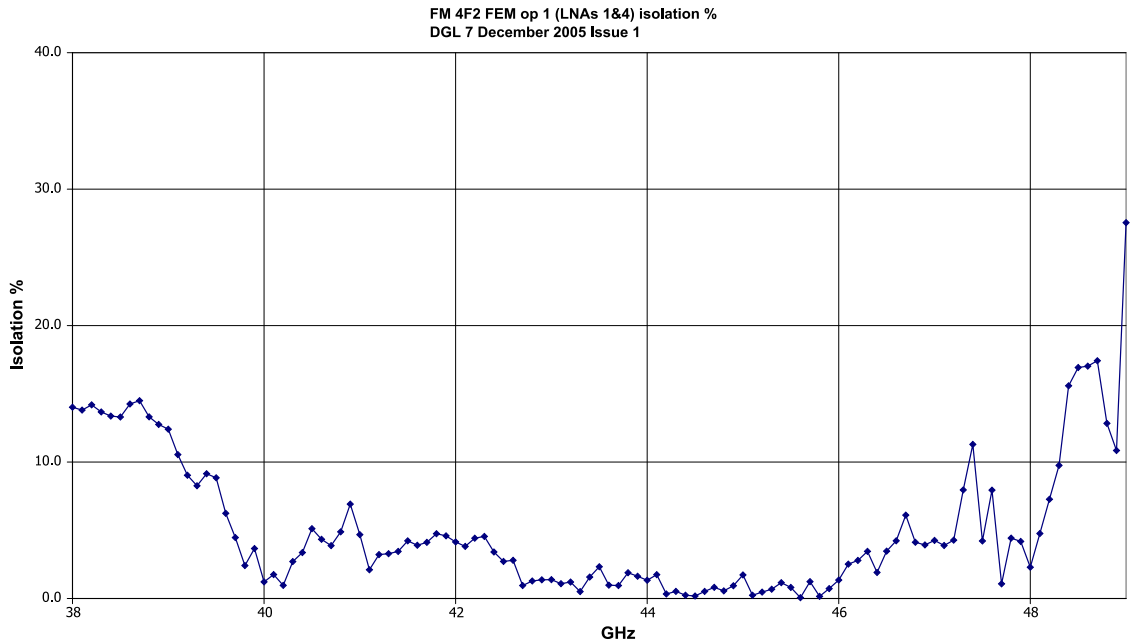}
			\caption{Isolation across the Planck band for FEM 4F2, output 1.  This is measuring the isolation between the channel containing LNA 4F2A1 and the channel containing LNA 4F2A4.}
			\label{fig:isolation4F2}
			\end{figure}

			The isolation was calculated in this way at all the measured points across the Planck band for each FEM, and the results shown in Table \ref{tbl:30flightperformance} and Table \ref{tbl:44flightperformance} are the band averages. Figure \ref{fig:isolation4F2} shows an example of the isolation variation with frequency for the FEM 4F2.  Typically the isolation was found to be well below the 10\% requirement at all points in the Planck band.  These results are obtained without a BEM, and refer to the FEM alone.

		\subsubsection{ Radiometer isolation including phase switch balance}
		\label{sec:radiometerisolation}

			The isolation between channels as seen in the radiometer configuration can also be obtained from the radiometer time-ordered data streams if the data from each radiometer are recorded separately (without double differencing, see Section \ref{sec:operationmode}).  Comparison of the output levels before and after the temperature rise give measurements of the total isolation, including the phase switch amplitude mismatch.

			\begin{figure}
			\centering
			\includegraphics[angle=0,width=15 cm]{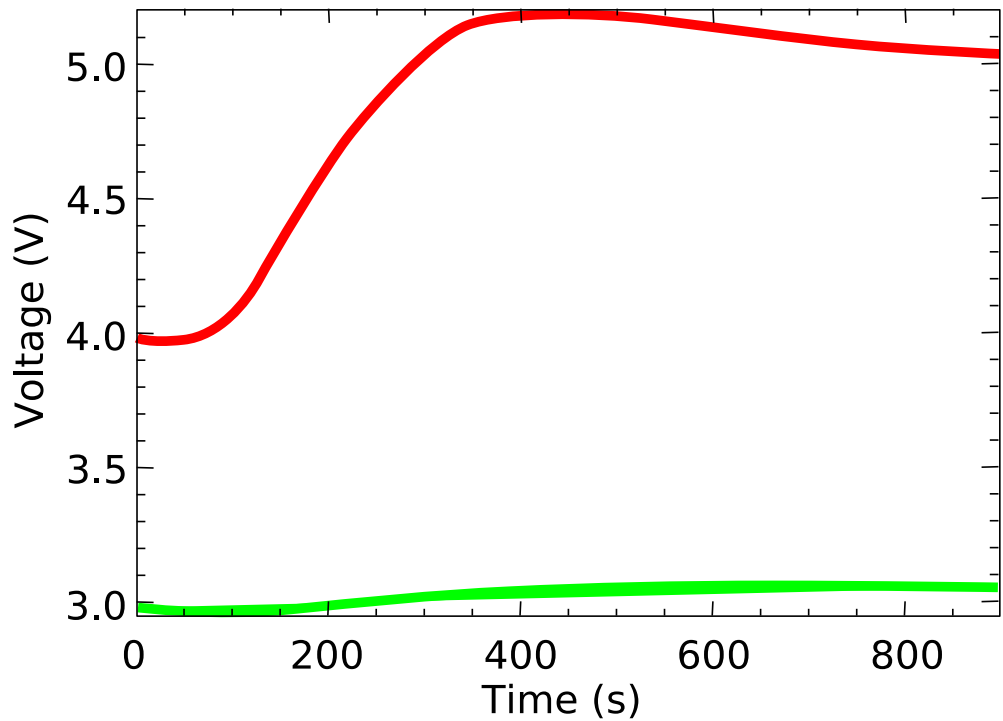}
			\caption{Run 6, time series ramp, 15 minutes data stream taken with FEM 3F2 and the two representative BEMs. Dark (red) line refers to temperature ramp and the light (green) line shows the cross-talk sigma}
			\label{fig:run6}
			\end{figure}

			Figure \ref{fig:run6} shows a 15 minute run taken with FEM 3F2 using the radiometer containing amplifiers 3F2A1 and 3F2A4 while the temperature on the variable load was ramped from 28.7 to 40~K. The time taken to stabilise the system after this temperature change is such that equilibrium has probably not quite been reached, but assuming the level at the end of the run is close to the equilibrium level, the isolation, and isolation corrected for parasitic heating of the ``fixed'' load as described in Section \ref{sec:isolationfreq} (Equation 3) are as shown in Tables \ref{tbl:tempramp} and \ref{tbl:isolation3F2}. The long time (several 100~s) the signal takes to reach its final value is caused by a combination of time constants of the different components which make the FEM, BEM, and data acquisition electronics. The isolation is well within the requirement of 10\%, and also within the goal figure of 5\% for both channels (both detectors).  The results are similar for all channels, at both frequencies. The isolation corrected for parasitic heating is compared in the next line with the uncorrected isolation determined in \ref{sec:isolationfreq}, which is slightly higher. The final line gives the value of the radiometer calibration constant (see Section \ref{sec:linearity}) which was used in the correction for the parasitic heating.

			\begin{table}
			\caption[]{\label{tbl:tempramp} Temperature ramp}
			\begin{center}
			\begin{tabular}{ l c c c c } \hline \hline
			~ &  & Thot (K) & Tcold (K) & Tfem (K) \\ \hline
			\multicolumn{2}{l}{Low end of ramp} & 28.7 & 17.8 & 17.9 \\
			\multicolumn{2}{l}{High end of ramp} & 40 & 18.6 & 18.7 \\ \hline
			\end{tabular}
			\end{center}
			\end{table}

			\begin{table}
			\caption[]{\label{tbl:isolation3F2} Isolation for FEM 3F2 as determined from radiometer ramp data. Radiometer data run with QM representative BEMs.}
			\begin{center}
			\begin{tabular}{ l c c c c } \hline \hline
			~ & \multicolumn{2}{c}{Detector 1} & \multicolumn{2}{c}{Detector 2} \\
			Channel & 0 & 1 & 2 & 3 \\ \hline
			Mean before ramp (V) & 2.95 & 3.97 & 3.95 & 2.97 \\ 
			Mean after ramp (V) & 3.05 & 5.14 & 5.13 & 3.08 \\
			Isolation \% & \multicolumn{2}{c}{8.15} & \multicolumn{2}{c}{8.69} \\
			Isolation \% corrected & \multicolumn{2}{c}{1.4} & \multicolumn{2}{c}{2.2} \\ 
			for parasitic heating &  & &  & \\
			Mean isolation as & \multicolumn{2}{c}{3.3} & \multicolumn{2}{c}{3.2} \\
			measured across band &  & & & \\
			Cal const, V/K & \multicolumn{2}{c}{0.104} & \multicolumn{2}{c}{0.104} \\ \hline 
			\end{tabular}
			\end{center}
			\end{table}

	\subsection{Radiometer tests, FEM+BEM}

		The results of some radiometer tests have already been mentioned.  However, before completing the discussion of the FEM testing in the radiometer configuration, it is useful to describe some details of the radiometer set up and operation mode

		\subsubsection{ Phase switch tuning procedure  }

			Before running the radiometer tests, it was necessary to balance the phase switches to ensure optimum performance. 

\begin{figure}
			\centering
			\includegraphics[angle=0,width=10 cm]{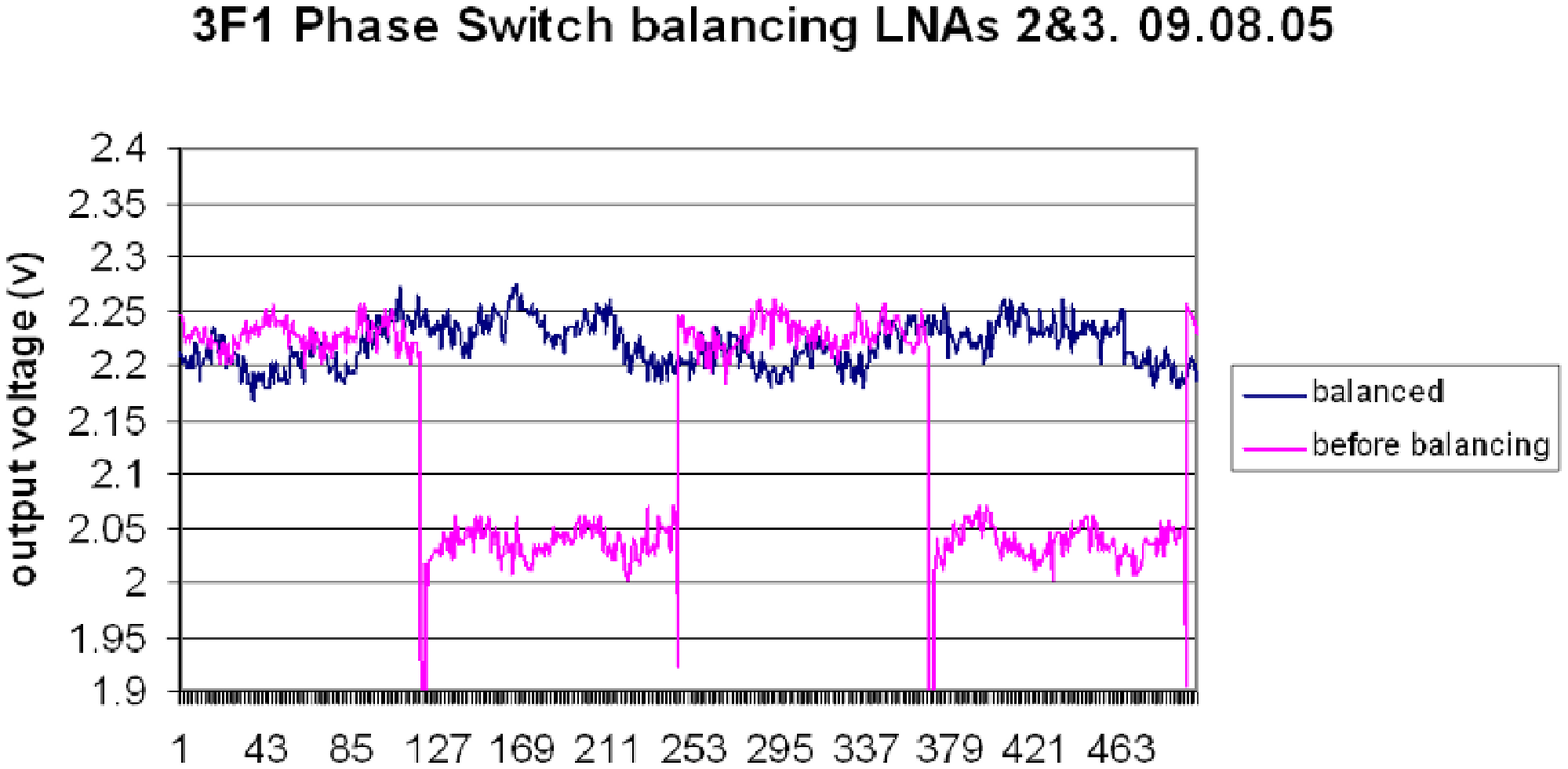}
			\caption{Oscilloscope trace of BEM output during phase switch balancing of FEM 3F1}
			\label{fig:oscilloscope}
			\end{figure}

			The balancing procedure is essentially a matching of phases through the two radiometer legs, and for this reason, the full radiometer chain with BEMs is required.  For this test, both sky and reference load are at the same temperature. The LNA in one arm of the radiometer is turned off, and the other arm is operated in switched mode. The FEM output is monitored on an oscilloscope, and the phase switch bias on the working arm is tuned until the amplitude in the square wave seen at the output of the BEM is zero, as illustrated in Figure \ref{fig:oscilloscope}. The procedure is then repeated for the other arm of the radiometer, and repeated iteratively if necessary until both outputs show no steps. This ensures that no offsets are introduced by a mistuning of the phase switches.

			In some cases, this process was found not to give any improvement in noise temperature once the amplifier pairs had been physically well matched.  Because the physical matching has been done with the phase switches in the standard operating configuration, the radiometer has essentially already been balanced, and any further adjustment tends to unbalance it.

		\subsubsection{ Noise Floor Verification}

			The white noise level sets the fundamental limit on the signal which can be measured, and the ratio $\Delta V/V$, where $\Delta V$ is the white noise level in $V/\sqrt{Hz}$, and $V$ is the average voltage level, gives the primary measurement of the sensitivity of the radiometer, independent of any detector offset. It is therefore essential to know that the measured white noise level is that of the radiometer itself, and not that contributed by the back end data acquisition electronics (DAE). This was confirmed by running the radiometer with the BEM on and off.

		\subsubsection{Operation mode - Differencing across the detector diodes}
		\label{sec:operationmode}

			The preceding tests on the FEM alone were all performed without switching. In order to test the performance likely in the full radiometer, in particular to determine the 1/f knee frequency, it was necessary to operate the FEM in switched mode. Strictly only one phase switch is needed in each radiometer, but to make the two arms of the radiometer as similar as possible for reasons of phase matching, a phase switch has been included in each arm (see Figure \ref{fig:blockdiagram}). During operation, only one phase switch is used, the other being fixed. This provides some redundancy in case of failure of a phase switch.

\begin{figure}
			\centering
			\includegraphics[angle=0,width=8 cm]{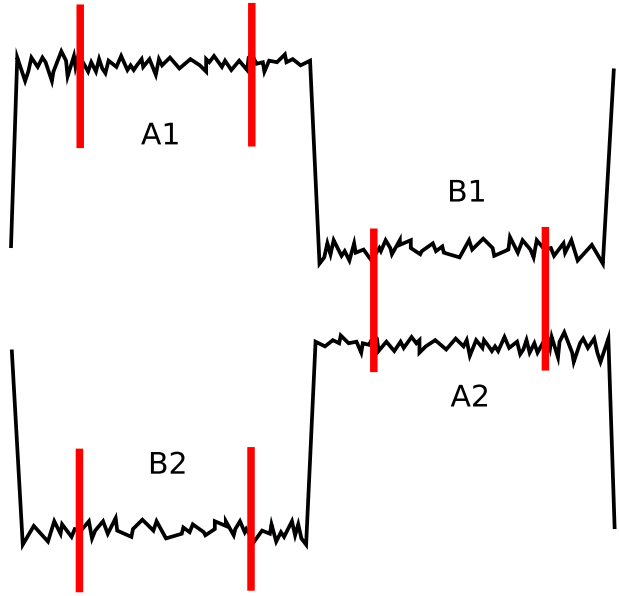}
			\caption{Illustration of a short section of radiometer output from the two radiometers in a single FEM. The red bars indicate the regions where the data are discarded at the beginning and the ends of the square waves. }
			\label{fig:radiometeroutput}
			\end{figure}

			Three data streams are recorded -- a switched waveform from each channel of the radiometer and one `pure' switched waveform from the phase switch (i.e. 0--5~V). The pure waveform is used as a trigger for post data analysis. The data are averaged over the step, resampled at the switch frequency, and binned into odd and even phase switch states, yielding 4 data streams. Quantities such as calibration constant can be determined directly from the undifferenced data.  However, to determine the 1/f  knee frequency the data must be differenced. The fact that the gains in each channel are not exactly the same is taken into account by means of the gain modulation or ``r'' factor, which should be less than 1.  In the illustration in Figure \ref{fig:radiometeroutput}, the difference would be (A1-B1)r- (A2-B2), where A1/B1 represent the two switch states on one radiometer output and A2/B2 are the two switch states on the second radiometer output. This first difference, effectively between the sky and the reference loads, removes long term drifts due to the amplifiers in the BEM. The mean white noise level may be calculated for each detector from this differenced data. Differencing again across the two detectors, the equivalent of a double Dicke system, removes drifts due to drifts in the FEM amplifiers. Fourier transforming the double differenced data yields the amplitude spectrum from which the 1/f knee frequency can be determined.

	\subsection{ Radiometer linearity and calibration curve}
	\label{sec:linearity}

		The radiometer constant, or correspondence between the measured voltage and the radiometer system temperature in K,  can be calculated from the ramp data shown in Figure \ref{fig:run6}. Typical values of the radiometer constant are given in Table \ref{tbl:isolation3F2}. The beginning and end level of the resulting ramp in the time series yield an immediate $\Delta V$ value for the $\Delta T$ increment (see \ref{sec:radiometerisolation}).  Alternatively, the constant can be measured directly by changing the input temperature and recording the radiometer output voltage change.  This test has been described in detail in \citet[][this volume]{2009_LFI_cal_R9}.

	\subsection{ 1/f knee frequency}

		\subsubsection{ Gain Modulation factor}

			Planck LFI uses the pseudo-correlation radiometer design \citep{2000ApL&C..37..171B}, where the temperature of the reference load does not have to be the same as the sky temperature because the so called ``gain modulation factor'', $r$, is used to null the output signal. Nulling the output signal minimizes the sensitivity to RF gain fluctuations. However, care must be taken because there will be effects on error propagation and levels of noise from the two channels are no longer identical.

			\citet{2002A&A...391.1185S} give the following expression for the 1/f knee frequency $f_k$:

			\begin{equation}
			f_{k} =\frac{1}{2} C^{2} \beta \frac{[(T_\mathrm{sky}^\mathrm{hyb} +(1+A/C)T_\mathrm{n} )-r(T_\mathrm{ref}^\mathrm{hyb} +(1+A/C)T_\mathrm{n} )]^{2} }{(T_\mathrm{sky}^\mathrm{hyb} +T_\mathrm{n} )^{2} },
			\label{eqn:fk}
			\end{equation}

			Which is zero for $r=\frac{T_\mathrm{sky}^\mathrm{hyb} +(1+A/C)T_\mathrm{n} }{T_\mathrm{ref}^\mathrm{hyb} +(1+A/C)T_\mathrm{n} }$.

			When $A/C=1/2\sqrt{N_\mathrm{s} } << 1$, $r=r_{0}^{*}$ and we have the condition for null radiometric power, with $T_\mathrm{sky}^\mathrm{hyb} =rT_\mathrm{ref}^\mathrm{hyb} $.  Then the knee frequency expression can be simplified as:

			\begin{equation}
			f_k\propto (1-r)^2 (T_\mathrm{n} / (T_\mathrm{n} + T~_\mathrm{x}))^2 \beta.
			\label{eqn:fk2}
			\end{equation}

			The gain fluctuations will cancel out and the residual 1/f noise is dominated by the noise temperature fluctuations in the HEMTs.

			No 4-K load was available at JBO, and the sky load was only maintained at about 20~K. In order to simulate the correct channel imbalance for the tests, as expected in flight, it was necessary to calculate the appropriate temperature for the reference load and the r value to which this corresponded.

			The radiometer tests in the cryostats were run with the reference load temperatures derived by this method at both frequencies, to simulate the 1/f conditions likely to prevail on Planck.  Time series data were taken and transformed to yield the power spectrum and hence the measured 1/f knee frequency.

			For Planck, it is extremely important to have data with the lowest 1/f knee frequency possible.  Excessive 1/f noise would degrade the data by increasing the effective rms noise, decreasing the sensitivity, and increasing the uncertainty in the measurement of the power spectrum at low multipoles. The post-detection knee frequency needs to be significantly lower than the satellite rotation frequency ($f_{spin} \sim 0.017$~Hz), otherwise the resultant maps will require de-striping. Although at values of $f_{k} \le 0.1$~Hz  it is possible to apply destriping algorithms to control the increase in rms noise to within a few \% of the white noise \citep{2002A&A...387..356M}, it is clearly preferable to avoid such measures.

\begin{figure}
			\centering
			\includegraphics[angle=0,width=15 cm]{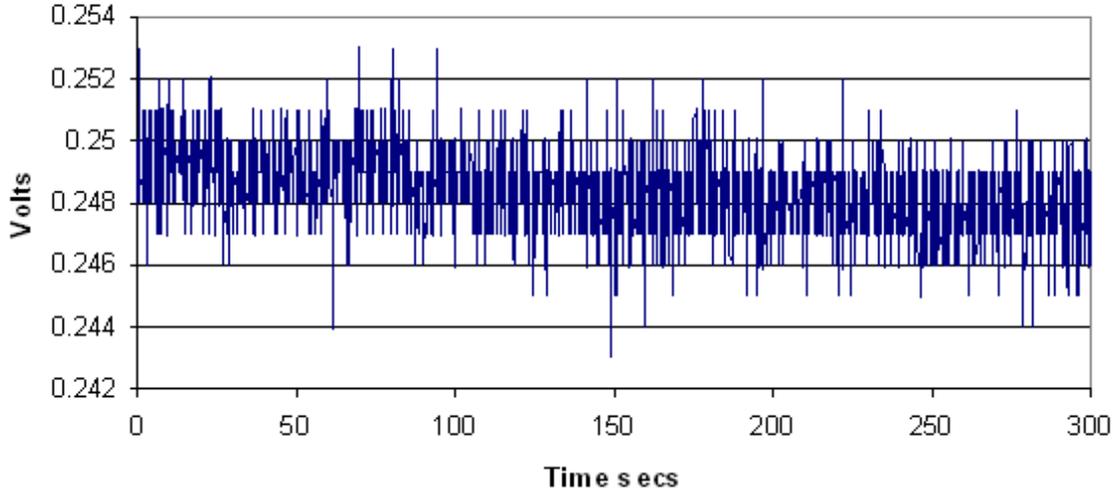}
			\caption{Example of time series}
			\label{fig:timeseries}
			\end{figure}

\begin{figure}
			\centering
			\includegraphics[angle=0,width=15 cm]{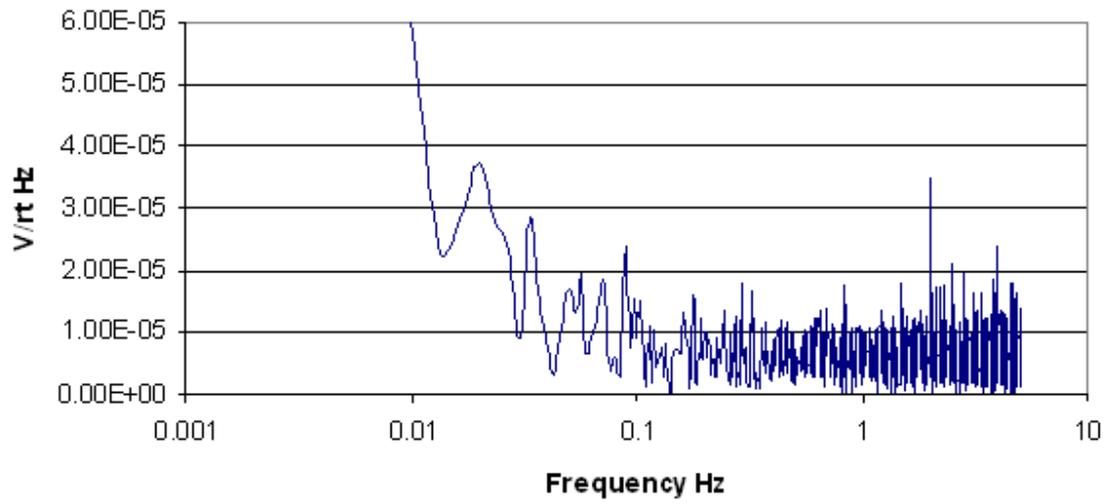}
			\caption{Frequency Spectrum, singly differenced data}
			\label{fig:frequencyspectrum}
			\end{figure}

\begin{figure}
			\centering
			\includegraphics[angle=0,width=15 cm]{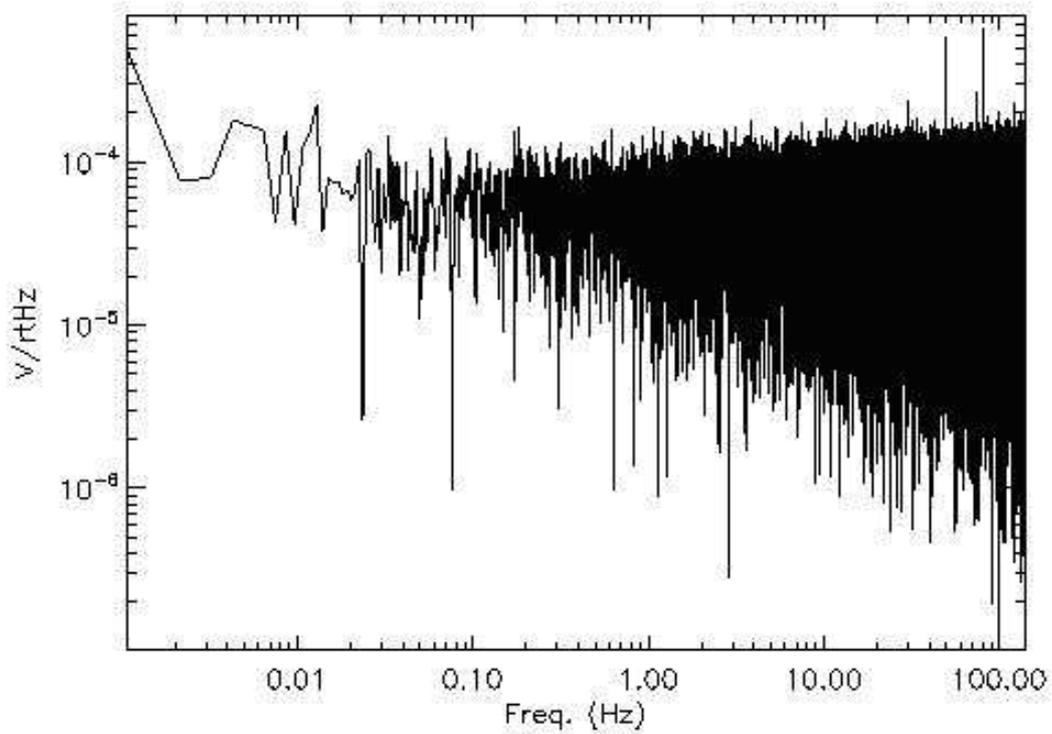}
			\caption{Power spectrum of the doubly differenced data for 4F2}
			\label{fig:powerspectrum}
			\end{figure}

			Figures \ref{fig:timeseries} to \ref{fig:powerspectrum} show respectively an example of a section of a typical time sequence, a power spectrum taken from a time sequence where the data have been differenced between the sky and reference load only (singly differenced) and an example of a power spectrum where the data  for a longer stream have been differenced also between the two detector diodes of the two output streams.

			The gain drift in the time sequence is clearly evident.  The first power spectrum shows a knee frequency of about 50~mHz, whereas the power spectrum of the doubly differenced data indicates a knee frequency of about 10~mHz.  It proved difficult to achieve consistency in these measurements in the laboratory set up at JBO, due to the presence of many other operations and frequencies on site. Hence the values quoted for the knee frequency in Section \ref{conclusions} are those obtained in the test campaign in Laben \citep[][this volume]{2009_LFI_cal_R2} where individual Radiometer Chain Assemblies (RCAs) were tested and then Radiometer Assembly Arrays (RAAs; the complete LFI) were tested.

		\subsubsection{Variation of 1/f knee frequency with gain modulation r}

			The gain modulation factor may vary during long integrations, and it may be necessary to update it during measurements.  In flight, it will be possible to determine the $r$ factor from the measured data by three different methods \citep[][]{2003A&A...410.1089M}. Assuming the offset balancing will be done in software, the r value can be calculated from radiometer data acquired in total power mode, i.e. before differencing.  The plan is to download about 15 minutes per day of total power mode data. The r factor can then be calculated from i) the average of the sky and reference loads, ii) the ratio of the sky and reference load standard deviations, and iii) by minimising the knee frequency determined from the final differenced data.

\begin{figure}
			\centering
			\includegraphics[angle=0,width=15 cm]{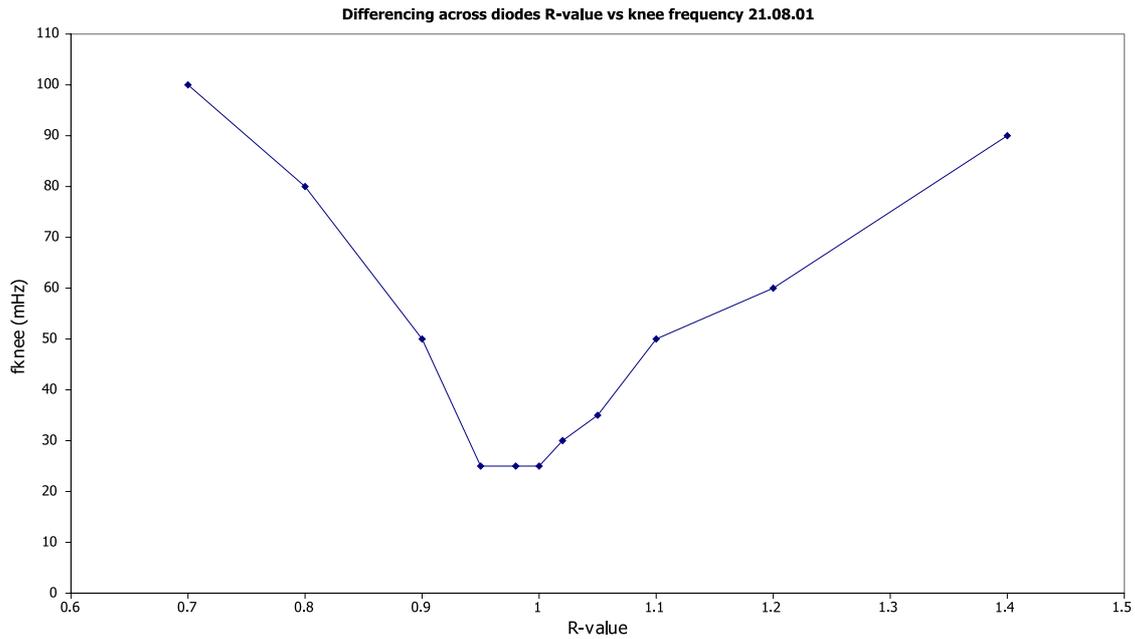}
			\caption{Variation in knee frequency with r-value}
			\label{fig:knee}
			\end{figure}

			An example of such a minimization is shown in Figure \ref{fig:knee}, where several different r values have been used, and the knee frequency resulting from each choice has been determined.

			This method is time consuming, so for online analysis, the other two methods mentioned will be preferable.

	\subsection{ Power consumption}

		The satellite as a whole has a stringent power budget because the cooling capacity of the cooler chains is limited. The power budget of each FEM was therefore carefully evaluated. The power dissipation per stage for each of the amplifiers was combined with a contribution for the phase switch and the 16 transistors used in the bias protection circuits to give the power dissipation for the whole FEM.  The results for each FEM are shown in Table \ref{tbl:powerdissipation}.  The total power dissipation summed over all five FEMs was within the requirements at 30 and 44~GHz.

		\begin{table}
		\caption[]{\label{tbl:powerdissipation} Power dissipation for all the flight FEMs}
		\begin{center}
		\begin{tabular}{ c c } \hline \hline
		FEM & Total Power (mW) \\ \hline 
		3F1 & 32.8 \\
		3F2 & 27.5 \\
		4F1 & 47.6 \\
		4F2 & 47.7 \\
		4F3 & 43.7 \\ \hline 
		\end{tabular}
		\end{center}
		\end{table}

	\subsection{Polarisation Isolation}

		Each radiometer chain assembly (RCA) containing one FEM, includes two radiometers, one for the ``E'' and one for the ``H'' linear polarizations, collected simultaneously for a single horn. Any leakage between any of the channels will be interpreted as an additional systematic polarized signal, so it is important that the cross polarization leakage, or polarization isolation, should be as low as possible. The mean cross polarization was found to be between -51 and -58~dB for all combinations. The leakage from one channel back into the sky port of the other channel is between -32 and -34~dB, which is very much less than is reflected back from the input itself (S11). These results ensure excellent polarisation isolation in the FEM for the purpose of CMB polarisation measurements. The polarisation properties of these radiometers will thus be controlled by the OMT properties.

\section{Conclusions}
\label{conclusions}

	We report in this paper the construction and testing of the FEMs at 30 and 44~GHz for the LFI in the  Planck mission \citep[][this volume]{2009_LFI}. The FEMs and BEMs, when combined, provide the linearly polarized radiometers at these frequencies. The specifications for these FEM units were agreed at the outset of the project \citep{2000ApL&C..37..171B}.  

	The LFI FEM parameters necessary to meet the science objectives at 30 and 44~GHz were given as requirements and goals and are summarised in Table \ref{tbl:summary} where they are compared with the values actually achieved.  The FEM units meet the requirements, within the measurement errors, for most parameters and in particular the noise temperature. The units come close to the more stringent goals in several parameters.

	\begin{table*}
	\caption[]{\label{tbl:summary} Summary of the FEM goals, requirements and mean achieved performances.}
	\begin{center}
	\begin{tabular}{ l c c c c } \hline \hline
	 & Centre & Goal & Requirement & Achieved \\
	 & Frequency & & & \\
	 & (GHz) & & & \\ \hline
	Gain, excluding phase switch & 30 & 33~dB & 30~dB (including phase switch) & 31.1~dB (mean) \\
	 insertion loss & 44 & 33~dB & 30~dB (including phase switch) & 32.3~dB \\
	Noise temperature of the FEMs & 30 & 6.1~K & 8.6~K & 8.9~K (mean across band) \\
	 & 44 & 10.4~K & 14.1~K & 15.6 (mean across band) \\
	Bandwidth & 30 & 6~GHz & 6~GHz & 5.7~GHz \\
	 & 44 & 8.8~GHz & 8.8~GHz & 7.3~GHz \\
	Isolation & 30 & $<$5\% & 10\% & $\sim$~4.0\% \\
	 & 44 & $<$5\% & 10\% & 4.1\% \\
	1/f knee frequency & 30 & 20~mHz & $<$50~mHz & $\sim$~28~mHz \\
	 & 44 & 20~mHz & $<$50~mHz & $\sim$~29~mHz \\
	Temperature stability requirements &  &  & $10\mu K Hz^{-1/2} >$10~mHz & \\
	 &  &  & $100\mu K Hz^{-1/2} <$10~mHz & \\ \hline
	\end{tabular}
	\end{center}
	\end{table*}

	Of particular note are the noise temperatures achieved; these along with the wide  bandwidths  are critical for the high sensitivity required for the Planck mission. Some LNAs within the FEMs (see Tables \ref{tbl:meanperformance30} and \ref{tbl:meanperformance44}) met the goals at 30~GHz and 44~GHz within the measurements errors and reached 3 and 5 times the theoretical quantum limit  respectively at the band centres. Furthermore, a range of tests showed that LNAs and FEMs achieved the stability levels required to meet the observing strategy of Planck.  In particular, the 1/f noise knee frequency $\le$29~mHz, close to the goal, met the conditions imposed by the 60 second rotation period of the spacecraft.

	The linear polarization performance of the FEMs exceeded the requirements of the mission.  The isolation between the E- and H- polarizations was measured to be between 51 and 58~dBs for the various FEMs. The LFI radiometers have very well determined position angle precision, being determined by the accuracy of the waveguide engineering.  The 30 and 44~GHz geometry is accurate to $\sim$~0.1\degr; the corresponding precision is $\sim$~1\degr in the HFI polarimeters. The temperature stability requirement values are given in Table \ref{tbl:summary}. The temperature accuracy is controlled by the HFI sensors and will be published in a later HFI paper. Planck will thus add these temperature values of the 4~K load to the differential measurements of LFI to give absolute temperature results.

	The final system temperatures will include a contribution from the CMB and a possible contribution from the spacecraft environment. All the FEMs passed their Test Review Board and were deliverd to ESA and shipped to and received in Laben, on schedule. They were then integrated into the Radiometer Chain Assembly and eventually in to the Radiometer Array Assembly of LFI at Laben.  They underwent further testing at each stage, and were finaly integrated to the spacecraft in Cannes and the whole instrument was cryogenically tested for the first time in CSL, Liege. \citep[see][]{2009_LFI_cal_M3,2009_LFI_cal_M4}

\section*{Acknowledgements}
	Planck is a project of the European Space Agency with instruments funded by ESA member states, and with special contributions from Denmark and NASA (USA). The Planck-LFI project is developed by an International Consortium lead by Italy and involving Canada, Finland, Germany, Norway, Spain, Switzerland, UK, and USA. We would also like to thank Keith Williams of UMIST Departement of Electronic and Eleactical Engineering (now the University of Manchester) for measuring device data on wafer, Helen Yates for contibutions to the  radiometer testing, Bert Fujiwara and Mary Wells of JPL for assistance with LNA fabrication and testing, Prof. J. Cooper and A.Ainul, of the Electrical Engineering department of the University of Manchester for the structural analysis, and STFC for funding the UK LFI programme.

\bibliographystyle{aa} 
\bibliography{davis} 

\end{document}